\documentclass{iopart}

\usepackage[pdftex]{graphicx}
\usepackage[pdftex]{epsfig}
\usepackage{epstopdf}
\usepackage{dcolumn}
\usepackage{bm}
\usepackage{color}

\newcommand{\units}[1]{\ensuremath{\mathrm{#1}}}
\newcommand{\amount}[2]{\ensuremath{#1\:\units{#2}}}

\begin{document}

\paper[Valley splitting in a Si/SiGe QPC]{Valley splitting in a Si/SiGe quantum point contact}

\author{L M McGuire, Mark Friesen, K A Slinker, S N Coppersmith and M A Eriksson}
\address{University of Wisconsin-Madison, Physics Department, 1150 University Ave, Madison, Wisconsin 53706, USA}

\begin{abstract}
We analyze transport data from a quantum point contact (QPC), fabricated on a modulation doped Si/SiGe heterostructure, to extract experimental estimates for the valley splitting. 
The experimental data are fit to a form derived from a valley coupling theory that
takes into account the fact that the quantum well is grown on a miscut substrate. 
The results of the fitting analysis are compared to results obtained by fitting to a simple
phenomenological form; both methods indicate that electrostatic and magnetic confinement enhance the valley splitting by reducing the lateral spatial extent of the electronic wavefunction.
  Consequently, the valley splitting can be much larger than the spin splitting for small magnetic fields.  We observe different valley splittings for the two lowest orbital modes of the QPC, supporting the notion that when steps are present at the quantum well interface, the spatial extent of the wavefunction plays a dominant role in determining the valley splitting.
\end{abstract}

\submitto{\NJP}

\section{Introduction}\label{chap:Intro}Quantum computing relies on the existence of a well defined two-level quantum system, or qubit, that is separated in energy from all other nearby levels. There are a number of experimental systems where this situation is realized, including NMR \cite{Vandersypen:2001p883}, optics \cite{Turchette:1995p4710}, trapped atoms and ions \cite{Sorensen:1999p1971}, and a number of solid state systems \cite{Averin:1998p659,Loss:1998p120,Kane:1998p133}.  Spin states in semiconductor quantum dots form a promising qubit candidate because their lifetimes can be quite long \cite{Elzerman:2004p431,Johnson:2005p925}.  Spin measurement and quantum gates have been performed in GaAs quantum dot qubits  \cite{Elzerman:2004p431,Johnson:2005p925,Ciorga:2000p16315,Petta:2005p161301,Petta:2005p2180,Koppens:2006p766,Nowack:2007p1430,Koppens:2008p236802,Reilly:2008p817,PioroLadriere:2008p776,Barthel:2009p160503}.  While still less developed than their III-V counterparts, significant progress has also occurred in silicon quantum dots in recent years \cite{Simmel:1999p10441,Rokhinson:2001p035321,Klein:2004p4047,Zhong:2005p1143,Slinker:2005p246,Sakr:2005p223104,Berer:2006p162112,Jones:2006p073106,Klein:2007p033103,Zimmerman:2007p033507,Simmons:2007p213103,Shaji:2008p540,Angus:2008p112103,Liu:2008p073310,Fuhrer:2009p707,Zwanenburg:2009p1071,Simmons:2009p3234,Nordberg:2009p115331,Nordberg:2009p202102}.

\bibliographystyle{unsrt}

The advantages of silicon as a host for spin qubits include
spin-zero nuclei that are naturally abundant and
small spin-orbit coupling \cite{Kane:1998p133,Friesen:2003p121301,Morello:2009p081307},
properties that favor
long spin lifetimes \cite{Feher:1959p1219,Wilson:1961p1068,Tyryshkin:2003p193207}.  However, silicon  also possesses nearly-degenerate conduction band minima, or valleys, that could interfere with qubit operations or open additional pathways for decoherence \cite{Tahan:2002p035314,Koiller:2001p027903,Boykin:2004p115,Boykin:2004p165325}.  In quantum wells, the valley degeneracy is lifted, but the problem of decoherence or interference still exists if the splitting is smaller than or comparable to the spin splitting \cite{Eriksson:2004p133}.

Several previous measurements of the valley splitting were made in laterally unconfined Si/SiGe two-dimensional electron gases (2DEGs).  In these experiments, the valley splitting was found to be much smaller than the spin splitting, and to depend strongly on the perpendicular magnetic field \cite{Weitz:1996p542,Koester:1997p384,Schumacher:1998p260,Wilde:2005p165429,Lai:2006p076805,Lai:2006p161301,Goswami:2007p41}.  Neither of these observations can be explained adequately without taking into account the presence of atomic steps at the quantum well interface \cite{Ando:1979p3089,Friesen:2006p202106}.  When steps are present, the valley splitting is expected to be larger for wavefunctions that cover fewer steps \cite{Goswami:2007p41,Friesen:2006p202106,Friesen:2007p115318,Kharche:2007p092109,Friesen:2009preprint}.  Controlled lateral confinement -- for example, in a quantum point contact or a quantum dot -- therefore provides a means to manipulate the valley splitting.

In this paper, we provide a detailed follow-up to a quantum point contact (QPC) experiment reported in \cite{Goswami:2007p41}.   
Both this paper and \cite{Goswami:2007p41}
analyze QPC conductivity data
to obtain the spectrum of energy levels and thus
the valley splitting. 
This method for determining valley splitting, which is
based on the early work of van Wees, requires a formula for the valley splitting as a function of magnetic field \cite{VanWees:1991p12431}.  
In \cite{Goswami:2007p41}, a simple functional form 
was used for this dependence.
In the present work, we analyze the QPC conductivity data
using a fitting form for the valley splitting that is theoretically derived, using a model for the valley splitting in a QPC geometry that includes interfacial steps as well as electrostatic and magnetic confinement.  The theory yields fitting forms for the valley splitting for each of the magnetic orbitals, known as Fock-Darwin (FD) levels.  Because multiple FD levels are fit using the same parameters, the theoretically justified fitting form requires fewer parameters than the phenomenological form used in \cite{Goswami:2007p41}.  

We compare the results from the different analyses by carrying
out a detailed statistical analysis to assess the error bars and the range of uncertainty for the fit parameters.  
Even though the asymptotic behaviors of the fitting forms
at large magnetic fields are different,
 they yield similar values for the valley splitting in the regime where the experimental data are dense.  
For both methods, the valley splitting in a laterally confined quantum device is relatively large (a substantial fraction of a millivolt),
consistent with the hypothesis that lateral confinement increases the valley
splitting by reducing the number of interfacial steps covered by the electronic
wavefunction.

The paper is organized as follows.  In section~\ref{chap:data}, we review the experimental details of the conductance measurements in a QPC.  In section~\ref{chap:valley}, we discuss the valley splitting theory and the fitting forms used in our analysis.  In section~\ref{chap:analysis}, we present our method for extracting the valley splitting.  In sections~\ref{sec:results} and \ref{sec:discussion}, we present and discuss the results of the fitting analysis.  In the appendix, we provide details of the theory of valley splitting in a QPC in the presence of a stepped interface.  

\section{Experiment and analysis}\label{chap:data}\subsection{Experiment}\label{sec:experiment}
\begin{figure}[t]
 \begin{center}  \includegraphics[width=4.5in]{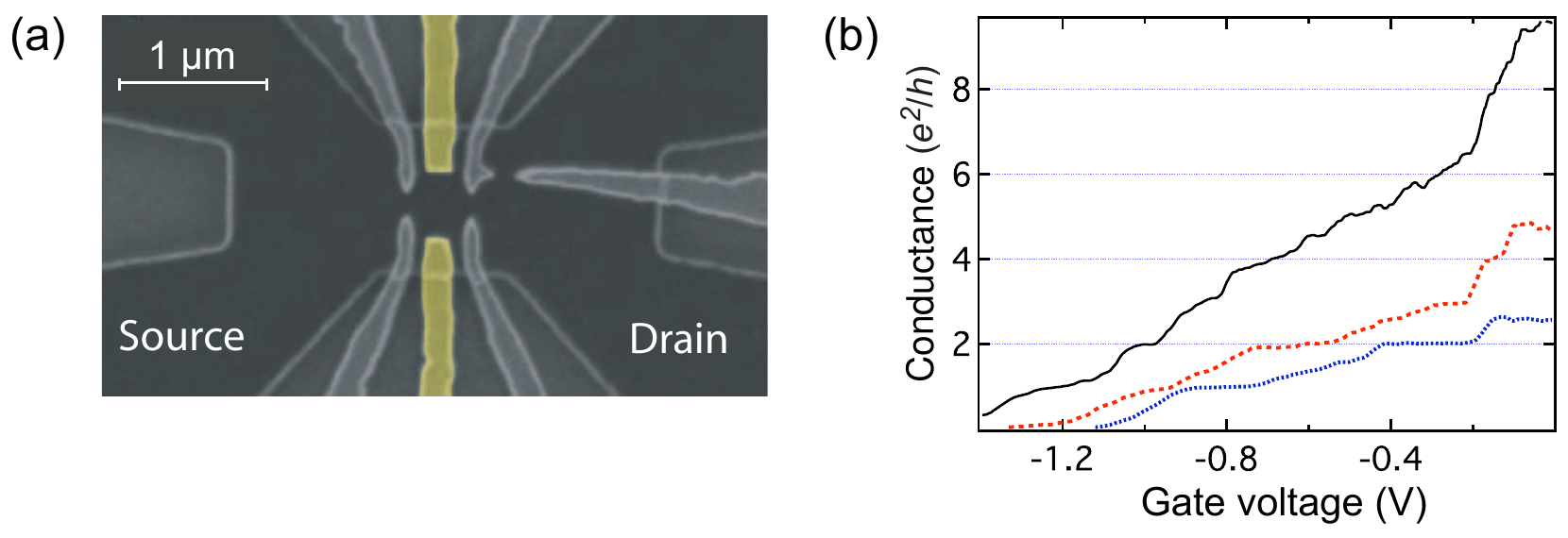}  \end{center}
 \caption{
 (a) Scanning electron micrograph of a top-gated Si/SiGe device, operated as a quantum point contact (QPC).  In this work the QPC was formed between the highlighted gates.  The 2-point conductance was measured between the source and the drain.  (b) Experimentally measured conductance through the QPC as a function of gate voltage, for constant magnetic fields of 2~T (solid black), 4~T (dashed red), and 6~T (dotted blue).}
\label{fig:scans}
\end{figure}

The QPC from \cite{Goswami:2007p41} was fabricated in a Si/SiGe modulation-doped quantum well grown at IBM-Watson in the mid-1990s by ultra-high vacuum chemical vapor deposition on a $p$-type Si substrate at a growth temperature of $560~^\circ$C.  The method is based on the approach reported in \cite{Ismail:1995p1077}.  The following growth profile for our sample wafer was originally reported in \cite{Koester:1997p384}  (referred to therein as wafer MOD42):  1~$\mu$m of SiGe, step-graded from Si to Si$_{0.75}$Ge$_{0.25}$; 1~$\mu$m of Si$_{0.8}$Ge$_{0.2}$; 10~nm of Si (the quantum well); 15~nm of Si$_{0.75}$Ge$_{0.25}$; 10~nm of Si$_{0.75}$Ge$_{0.25}$, doped with phosphorus at the density $N_d =(2$-$3)\times 10^{18}\, {\rm cm}^{-3}$; a 4~nm Si cap layer with $N_d =(4$-$10)\times 10^{18}\, {\rm cm}^{-3}$.  The resulting two-dimensional electron gas density was \amount{5.7 \times 10^{11}}{cm^{-2}} and had a mobility of 200,000~cm$^{2}$/Vs.  The nanostructured gates (palladium, deposited by electron beam evaporation) were formed on a small mesa etched into the heterostructure.  The mesa was defined lithographically, and the etching was accomplished with a CF$_4$ reactive ion etch. The etch depth was $\sim 70$~nm.  An electron micrograph of the device is  shown in figure~\ref{fig:scans}(a).  The yellow-highlighted palladium Schottky top-gates were used to form a quantum point contact by applying negative voltages with respect to the underlying electron gas.    A source-drain bias of 10~$\mu$V at 112~Hz was applied, and the 2-point conductance was measured as the gate voltage was swept from 0 to -1.4~V in steps of -0.002~V.  A perpendicular magnetic field was stepped from 0 to 8~T in increments of 0.1~T.  

Figure~\ref{fig:scans}(b) shows typical conductance scans as a function of gate voltage.  The measured conductance was corrected for a series of background resistances, including ohmic contacts, the 2DEG far away from the QPC, and resistors in the filter stage, so that the conductance plateaus lined up at multiples of $e^2/h$.   The lowest plateau occurs at $e^2/h$, indicating that the valley and spin degeneracies are completely lifted in this regime.  The observation of a plateau at $e^2/h$ in a Si/SiGe QPC was previously reported by T\"obben \etal \cite{Tobben:1995p711} and Scappucci \etal \cite{Scappucci:2006p035321}, and in Si/SiO$_2$ inversion layers by Wang \etal \cite{Wang:1992p12873}.

\subsection{Spectrocopic method}\label{sec:spectroscopy}
Van Wees \etal \cite{VanWees:1991p12431} have developed a spectroscopy technique for QPCs, using the fact that, at finite magnetic fields, the conductance is quantized into multiples of $e^2/h$.  Transitions between conductance plateaus occur whenever the minimum energy of a one-dimensional subband crosses the Fermi level.  Experimentally, the position of each transition can be mapped out as a function of gate voltage and magnetic field.  Theoretically, these transitions can be further mapped onto energy splittings, using a model for the QPC subbands.

An appropriate theoretical expression for the quantum levels in a QPC must take into account spin and orbital effects.  In Si, our model must also include valley effects, as described in \ref{sec:VStheory}.  Treating the valley coupling perturbatively, we can characterize a single non-degenerate subband by three quantum numbers: the transverse confinement mode, or FD level ($n$), and the spin ($s$) and valley ($v$) quantum numbers.  Here, $n$ can be any non-negative integer, and $s$ and $v$ can take the values $\pm1$.  The energy for subband ($n,s,v$) is then given by 
\begin{eqnarray}
\label{eq:En1}
E_{n,s,v}&=&\left(n+\frac{1}{2}\right)\sqrt{(\hbar\omega_{\rm c})^2+(\hbar\omega_0)^2}+ \frac{1}{2}sg\mu_BB\\
&&\phantom{\left(n\right)\sqrt{\hbar\omega_{\rm c}^2}}+ \frac{1}{2}v\Delta_n(V_{\rm g},B)+eV_0+\frac{\hbar^2k^2}{2m^*},\nonumber
\end{eqnarray}
where $\hbar \omega_0$ and $\hbar \omega_{\rm c}$ are the electrostatic and magnetic confinement energies, respectively, the minimum energy of the confinement potential $eV_0$ is determined by the gate voltage $V_{\rm g}$, $g\mu_BB$ is the Zeeman energy, and $\hbar^2k^2/2m^*$ is the kinetic contribution due to electron motion along the channel.  A theoretical expression for the valley splitting $\Delta_n$ in orbital $n$ is derived in the appendix.

\subsection{Data fitting window}\label{sec:datarestrictions}
Our spectroscopy method relies on the presence of a well-formed QPC, which in turn requires full depletion underneath the Schottky gates.
For this reason, we use only data for which we know a quantum point contact is formed in the channel.  For gate voltages $-0.3<V_{\rm g}<0$~V, the conductance is dominated by the quantum Hall effect away from the QPC constriction.  For gate voltages $V_{\rm g}< -0.3$~V  there is a clear dependence of the conductance on both gate voltage and magnetic field. Therefore we restrict our data to the range between -0.3 and -1.4~V, the latter being the most negative gate voltage for which data was acquired.  We also restrict the data to magnetic fields greater than 1 T, because the uncertainty in the location of the conductance transitions becomes large when $B<1$~T.  See sections~\ref{sec:transmain} and \ref{sec:uncertainty} for details.

\section{Valleys in silicon/silicon-germanium heterostructures}\label{chap:valley}
Bulk silicon has six degenerate valley minima along the equivalent [001] directions near the Brillouin zone boundary at $k_0\simeq0.82(2\pi/a)$, where $a$ is the lattice constant of the silicon unit cell.  However, silicon grown on a relaxed Si$_x$Ge$_{1-x}$ virtual substrate is strained.  This effect partially lifts the sixfold degeneracy by causing the four in-plane valleys to rise in energy and the two out-of-plane valleys to fall in energy.  The resulting splitting is on the order of 200~meV for typical structures \cite{Schaffler:1997p1515}.  The anisotropic effective mass effectively enhances this splitting by 15-20~meV, due to confinement in a quantum well, similar to the effect in inversion layers \cite{Ando:1982p437}.  The lifting of the remaining two-fold degeneracy of the $z$-valleys is the valley splitting of interest in this paper.  It is caused by an abrupt confinement barrier at the quantum well interface.  The resulting valley eigenstates involve combinations of $k$-states from the two $z$-valleys.  The valley splitting arises from the potential energy difference in these two states \cite{Friesen:2007p115318}.

\begin{figure}[t] 
 \begin{center} \includegraphics[width=5in,keepaspectratio]{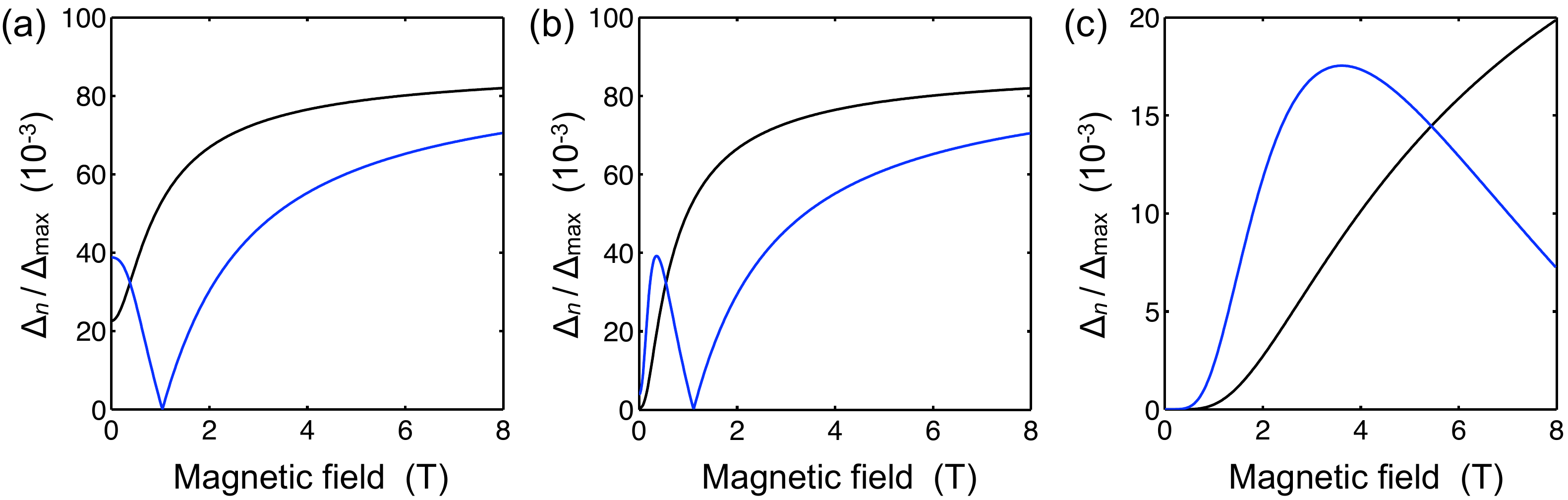}  \end{center}  
  \caption{
Scaled (dimensionless) valley splitting in a QPC, plotted as a function of magnetic field.
The plots show the valley splitting for the two lowest FD levels,  $\Delta_0$ (black) and
$\Delta_1$ (blue).  
(a) $L_y=40$~nm, $\theta =0.24^\circ$,
(b) $L_y=80$~nm, $\theta =0.24^\circ$,
(c) $L_y=40$~nm, $\theta =0.75^\circ$.
Note that $\theta =0.24^\circ$ is consistent with fitting results obtained in this paper.  For all three panels, we take $L_x=100$~nm and $\varphi = \pi/4$.  }
  \label{fig:VSQPC}
\end{figure}

Early theoretical work predicted a large valley splitting in the range 0.1-1~meV, for both Si/SiGe and Si/SiO$_2$ heterostructures \cite{Sham:1979p734,Ohkawa:1977p907,Ohkawa:1978p69}.  These theories assumed a perfectly flat quantum well with smooth interfaces, at zero magnetic field, and they ignored electron-electron interactions.  Later, numerical methods were applied under similar assumptions, confirming the previous estimates \cite{Boykin:2004p115,Boykin:2004p165325,Nestoklon:2008p155328}.  In the late 1990s, breakthroughs in the growth of Si/SiGe heterostructures made it possible to study new aspects of spin and valley physics. Many experiments were performed on Hall bars, and reported valley splittings that were much smaller than the predicted values \cite{Weitz:1996p542,Koester:1997p384,Schumacher:1998p260,Wilde:2005p165429,Lai:2006p076805,Lai:2006p161301,Goswami:2007p41}.  At zero field, most experiments suggested a valley splitting at least two orders of magnitude smaller than the theoretical estimates.  (Note that an anomalously large valley splitting has also been reported for a Si/SiO$_2$ quantum well \cite{Takashina:2006p236801}.  While small enhancements of the valley splitting have previously been attributed to many-body effects \cite{Khrapai:2003p113305}, the enormous enhancement reported in \cite{Takashina:2006p236801} has not been explained theoretically.)

Such unexpectedly small valley splitting measurements have been attributed to the presence of steps at the quantum well interface \cite{Goswami:2007p41,Friesen:2006p202106,Friesen:2007p115318,Kharche:2007p092109}, which cause a type of destructive interference that suppresses the valley splitting below its theoretical maximum.  Electron wave functions that cover many steps experience a large suppression in valley splitting.  Lateral confinement of the electron wavefunction decreases the destructive interference by reducing the number of steps that are covered.

In a QPC, the electron wavefunction is electrostatically confined by the top-gates and is therefore expected to have a larger valley splitting than a Hall bar, even in the absence of a magnetic field.  Applying a perpendicular magnetic field to the QPC should further enhance valley splitting, due to magnetic confinement over the length scale
\begin{equation}
\ell_B=\sqrt{\frac{\hbar}{eB_\perp}}.
\end{equation}

In \ref{sec:VStheory}, we develop a theory of valley splitting in a QPC, taking into account both electrostatic and magnetic confinement.  The theory provides fitting forms for the valley splitting in the first two FD levels:
\begin{eqnarray}
\Delta_0&=& \tilde{\Gamma} e^{-b}, \label{eq:D0theory} \\
\Delta_1&=& \tilde{\Gamma}  e^{-b}|1-2b|. \label{eq:D1theory}
\end{eqnarray}
Here,
\begin{eqnarray}
\tilde{\Gamma}&=&\frac{\Delta_{{\rm max}}}{\sqrt{\pi}}\Bigg|\frac{\sin(k_0L_x\theta \cos \varphi)}{k_0L_x\theta \cos \varphi}\Bigg|\label{eq:theory1}, \\
b &=& \frac{\tilde{A}}{\sqrt{(\hbar \omega_{\rm c})^2 + (\hbar \omega_0)^2}} , \label{eq:theory0}
\end{eqnarray}
and 
\begin{equation}
\tilde{A}= \frac{(k_0\theta \sin \varphi)^2}{m^*e/\hbar^2} \label{eq:Atilde} ,
\end{equation}
where $\Delta_{\rm max}$ is the theoretical maximum for the valley splitting, obtained at a perfectly flat interface, $L_x$ is the length of the QPC, and $\theta$ and $\varphi$ describe the average tilt and orientation of the miscut substrate, respectively. 
In combination with (\ref{eq:En1}), these equations provide fitting forms for the QPC conductance transitions.
Note that the magnetic field dependence in (\ref{eq:D0theory}) and (\ref{eq:D1theory}) predicts an increase in the valley splitting with magnetic field, a phenomenon that has been observed repeatedly in Hall bars \cite{Wilde:2005p165429,Goswami:2007p41,Khrapai:2003p113305}.

This result expresses the valley splitting in a QPC as a function of (i) magnetic field, (ii) device parameters, and (iii) heterostructure parameters.  Before proceeding to the detailed fitting in section \ref{chap:analysis}, it is useful to discuss the main features of (\ref{eq:D0theory}) and (\ref{eq:D1theory}), in order to understand why these forms are physically reasonable and to see how some of the basic features can be observed in the data.  The full data analysis will be discussed in considerable detail in section~\ref{chap:analysis}.  Here, we simply note that transitions between QPC conductance plateaus, such as those in figure~\ref{fig:scans}(b), occur whenever the bottom of an energy subband $E_{n,s,v}$ crosses the Fermi level.  These transitions can be mapped out as a function gate voltage and magnetic field, as shown below in figure~\ref{fig:TransBslice}.  This transition map contains information about the various terms in (\ref{eq:En1}), including the valley splitting.  

In (\ref{eq:D0theory}) and (\ref{eq:D1theory}), the magnetic field and the device parameters appear in the valley splitting terms involving $\omega_c$ and $\omega_0$, respectively.  In (\ref{eq:D0theory}), the valley splitting of the lowest FD level increases monotonically with magnetic field.  Such behavior is manifested in figure~\ref{fig:TransBslice}(a) because the splitting between conductance transitions monotonically increases with field at high fields.  The heterostructure parameters $\theta$ and $\varphi$ determine the magnitude of the valley splitting in (\ref{eq:D0theory}) and (\ref{eq:D1theory}); the valley splitting may approach its theoretical maximum $\Delta_{\rm max}$ when $\theta$ is small.  For typical heterostructures, we expect to find $\Delta_{\rm max}$ of order 0.1-1~meV \cite{Friesen:2007p115318} (recall that the Zeeman energy splitting at 2~T is 232~$\mu$eV). In figure~\ref{fig:TransBslice}(a), the spacing between the conductance transitions at fields in the range 2-4~T is relatively homogeneous, indicating that the valley splitting and Zeeman energy splitting in that range are comparable.  Finally, we note that the exponential functions appearing in (\ref{eq:D0theory}) and (\ref{eq:D1theory}) are physically reasonable.  The parameter $b$, which appears in the argument, is a positive quantity that approaches 0 for large magnetic fields or strong electrostatic confinement, causing the valley splitting to saturate.  Saturation occurs because, at strong enough confinement, further lateral squeezing of the electron wavefunction produces diminishing returns in enhancement of the valley splitting.  In combination with small $\theta$, strong confinement enables the valley splitting to approach its maximum value, as we observe. While necessarily general, these observations indicate that (\ref{eq:D0theory}) and (\ref{eq:D1theory}) make some simple predictions that can be understood without detailed fitting.  To make further progress, a statistical analysis is essential, and we report the results of such an approach below.

In order to make contact with our previous work, we report in parallel an analysis of a set of equations for the valley splitting that were used in \cite{Goswami:2007p41}.  These are purely phenomenological and are given by
\begin{eqnarray}
\Delta_0(V_{\rm g},B)&=&\sqrt{\Delta_{V_{\rm g}0}^2+(\Delta_{B0}B)^2}\label{eq:pheno0},\\
\Delta_1(V_{\rm g},B)&=&\sqrt{\Delta_{V_{\rm g}1}^2+(\Delta_{B1}B)^2}\label{eq:pheno1}.
\end{eqnarray}
This form was chosen for its simplicity and its linear magnetic field behavior at large enough $B$, with slope $\Delta_{Bn}$, a feature that is consistent with Hall bar data.  The $\Delta_{V_{\rm g}n}$ terms, if nonzero, allow for a nonzero valley splitting at $B=0$.
In section~\ref{chap:analysis}, we assume that neither $\Delta_{V_{\rm g}n}$ nor $\Delta_{Bn}$ depend on magnetic field.  All the parameters may depend on gate voltage.  Because of the phenomenological nature of (\ref{eq:pheno0}) and (\ref{eq:pheno1}), the various terms may not all be statistically relevant.  We test for this possibility by considering all possible functional variations with the various parameters $\Delta_{V_{\rm g}n}$ and/or $\Delta_{Bn}$ set to zero.  The statistical relevance of each case can be assessed by means of an $F$-test, which weights the $\chi^2$ value for the fit by $1/\nu$, where $\nu = ({\rm number\ of \ data\ points})-({\rm number\ of\  fitting\ parameters})$ \cite{Bevington:2003}.  The details of the $F$-test as applied to the parameters in (\ref{eq:pheno0}) and (\ref{eq:pheno1}) are presented in \cite{McGuire:Thesis}.  Here, we simply note that the most statistically relevant model is the variation where $\Delta_{V_{\rm g1}}$ is set to zero ({\it i.e.}, excluded from the model), leaving $\Delta_{V_{\rm g}0}$, $\Delta_{B0}$, and $\Delta_{B1}$ as fitting parameters.  In the following discussion, all further references to the ``phenomenological fitting form" refer to this 3-parameter model.  

\begin{figure}[t]
 \begin{center}  \includegraphics[width=4.2in]{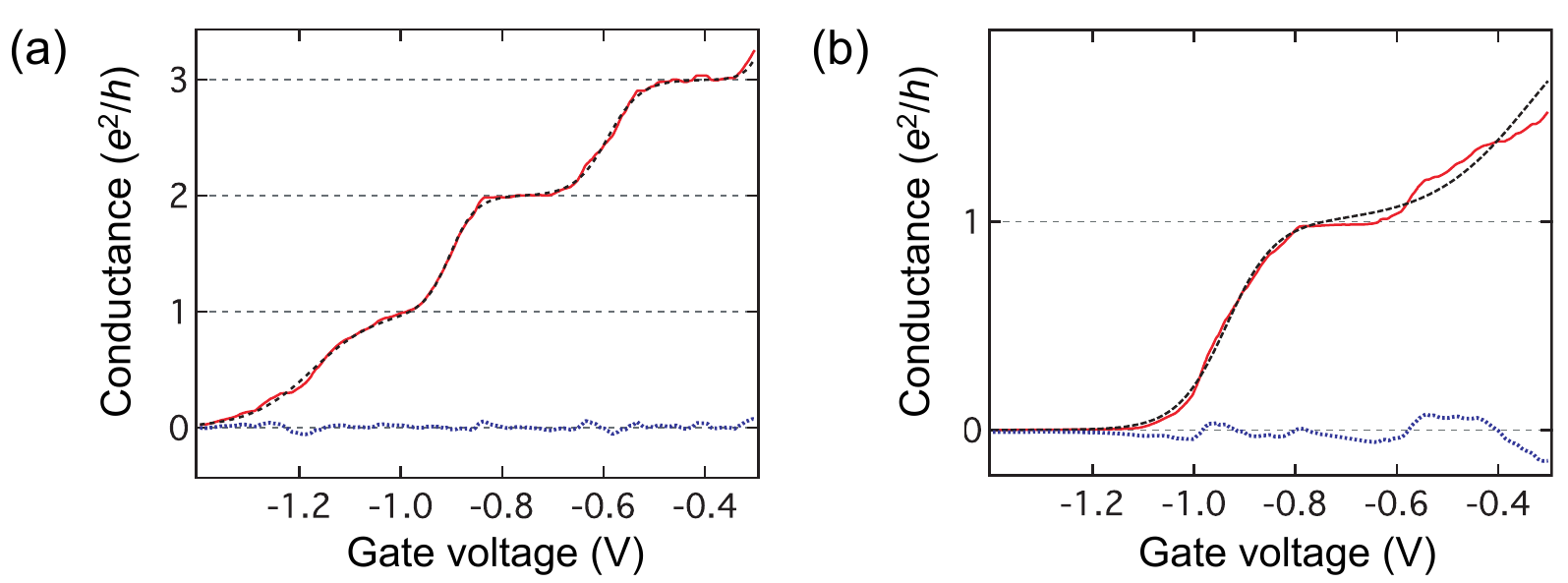}  \end{center}  
 \caption{
Experimentally measured
 conductance through a QPC, as a function of gate voltage.  Results are shown for magnetic fields (a) 3.5 T and (b) 7.4 T.  Solid, red curves:  conductance data. Dashed, black curves:  fits to Eq.~(\ref{eq:fermidirac}).  Dotted, blue curves:  fitting residuals.}
 \label{fig:FitswithRes}
 \end{figure}

It is important to note that the theoretical and phenomenological fitting forms have different asymptotic behaviors at high magnetic field.  At moderate field, in the regime where the data are dense (roughly $1T < B < 4$~T), we find that the separate analyses yield similar values for the valley splitting, as discussed in section~\ref{sec:discussion}.  Outside this field range, extrapolation of either model is uncertain.  We emphasize, however, that the phenomenological model grows linearly without bound at high fields, and is therefore physically unrealistic in such a limit.  In contrast, the theoretical model provides a physically reasonable saturation behavior.

\section{Data fitting analysis}\label{chap:analysis}
\subsection{Identifying the conductance transitions}\label{sec:transmain}
The first step in our spectroscopy analysis is to determine the locations of the conductance transitions in the QPC as a function of magnetic field and gate voltage.  A transition between conducting modes occurs whenever an energy minimum of a QPC subband crosses the Fermi level $E_{\rm F}$.  In (\ref{eq:En1}), this minimum corresponds to $k=0$.  
 
We consider a fixed magnetic field.  At finite temperature, the conductance increases gradually from plateau $(N-1)$ to plateau $N$, as shown in figure~\ref{fig:FitswithRes}.  Each step can be described by a Fermi-Dirac distribution function \cite{Beenakker:1991p1}, so that the conductance is given by
\begin{equation}
G=\frac{e^2}{h}\sum_i^N\frac{1}{1+{\rm exp}[(E_i-E_{\rm F})/k_{\rm B}T]}.
\end{equation}
The center of the $i^{\rm th}$ step corresponds to a particular gate voltage, such that $E_i(V_{\rm g})=E_{\rm F}$, where $i$ corresponds to a particular set of indices $(n,s,v)$.  The resulting transition gate voltage is denoted $V_{{\rm T}i}$. 

\begin{figure}[t]
 \begin{center}  \includegraphics[width=4.5in]{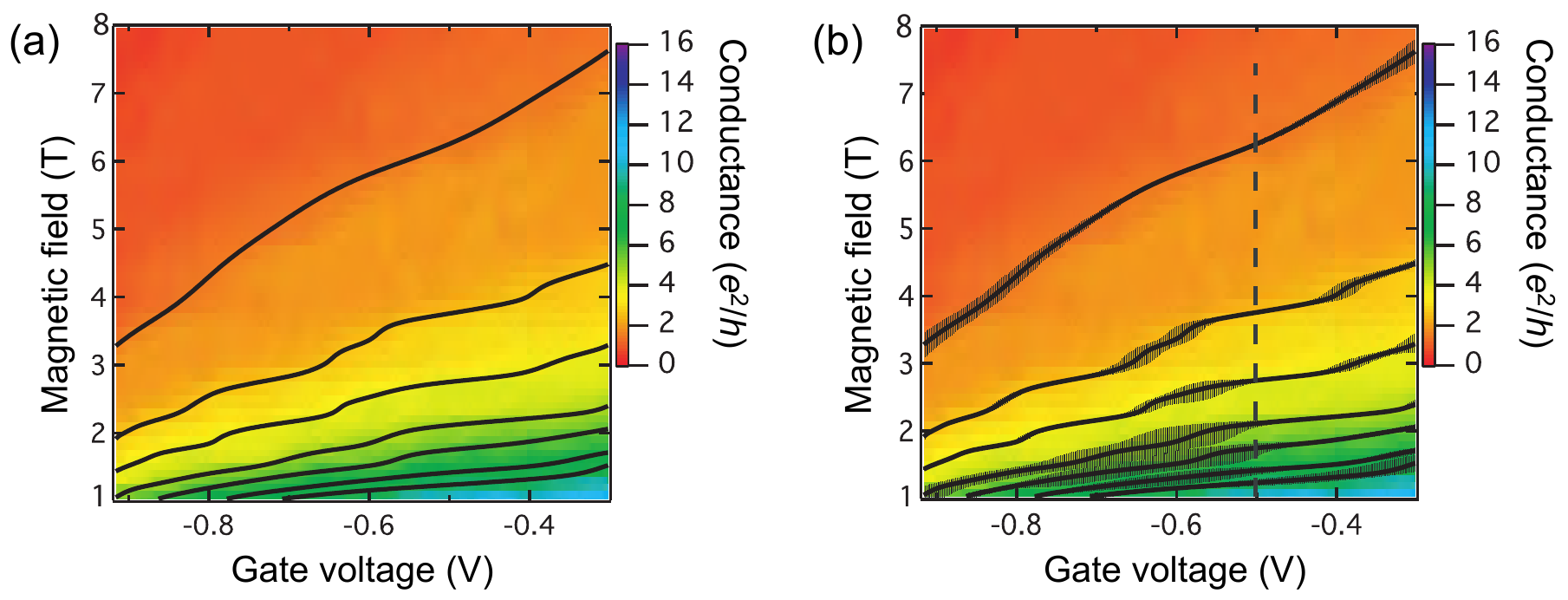}  \end{center}  
\caption{\label{fig:TransBslice}
(a)  Conductance as a function of gate voltage and magnetic field.  Solid black lines identify the conductance transitions with $(n,s,v)$ indices $(1,1,1)$, $(1,-1,1)$, $(1,1,-1)$, $(1,-1,-1)$, $(0,1,1)$, $(0,-1,1)$, and $(0,1,-1)$ (bottom to top). (b)  Same as (a), but including estimated 1-sigma error bars.  For a fixed gate voltage (\textit{e.g.}, the dashed line at $V_{\rm g}=-0.5$~V), the set of transitions $\{ B_{{\rm T}_{n,s,v}}\}$  defines a system of equations used to extract the valley splitting.}
\end{figure} 

Since (\ref{eq:En1}) describes an approximately linear relation between $V_{\rm g}$ and $E_i$, the conductance data may be fit to 
\begin{equation}\label{eq:fermidirac}
G=\frac{e^2}{h}\sum_{i=1}^{N}\frac{1}{1+{\rm exp}[-(V_{\rm g}-V_{{\rm T}_i})/W_i]} ,
\end{equation}
to extract the transition voltages.  Here, $W_i$ describes the broadening of the $i^{\rm th}$ transition in terms of gate voltage.  In this way, QPC conductance scans were fit for 71 fixed magnetic fields with $B>1$~T.  Typical results are shown in figure~\ref{fig:FitswithRes}.  For further analysis, it is convenient to express the conductance transitions in terms of gate voltage, rather than magnetic field.    Focusing on the lowest, $N\leq 8$, non-degenerate subbands, the data were therefore inverted to obtain the transition curves $B_{{\rm T}_i}(V_{\rm g})$ shown in figure~\ref{fig:TransBslice}.

\begin{figure}[t]
 \begin{center}  \includegraphics[width=4.in]{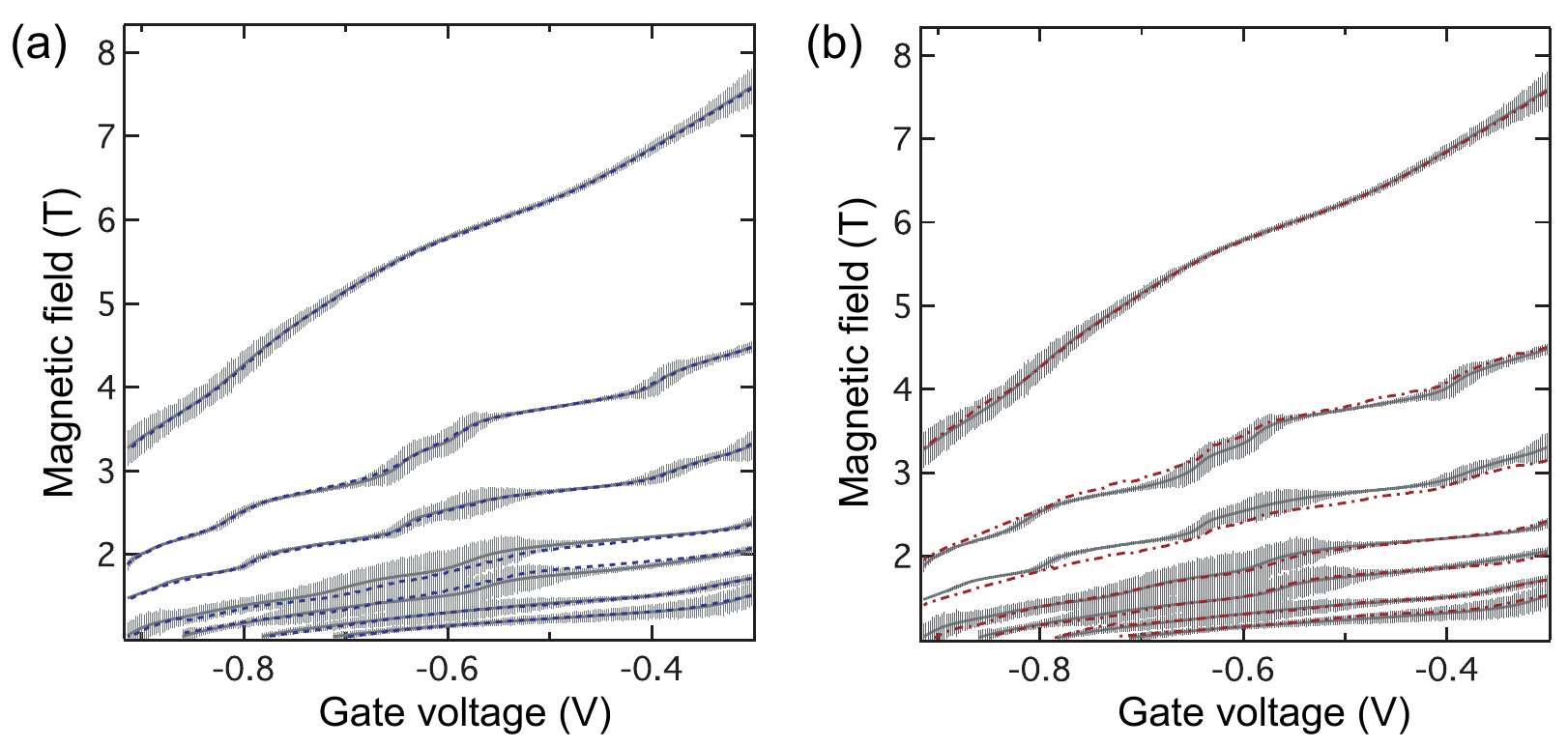}  \end{center}  
 \caption{ \label{fig:DNAall}
 Conductance transitions (solid gray lines) from figure~\ref{fig:TransBslice}.  Corresponding fits are obtained using (a) a phenomenological fitting form (dashed blue line), (b) a theoretical fitting form (dash-dot red line).  Note that the phenomenological form has a larger number of fitting parameters.  (See text.)}
\end{figure}

\subsection{Transition uncertainty}\label{sec:uncertainty} 
If there were no experimental uncertainty, the fitting routine described in section~\ref{sec:transmain} would give the precise location of each conductance transition.   In a real system, however, there is experimental noise which makes the transition location uncertain.  Additionally, the simple Fermi-Dirac function used in (\ref{eq:fermidirac}) does not account for electron correlations \cite{Reilly:2001p121311}, or interference effects as the electron enters or exits the QPC \cite{VanWees:1991p12431}.  These approximations, which have been left out of our analysis due to their complexity, could also, potentially, affect the location of the transition.

To obtain a quantitative estimate of the uncertainty in $B_{{\rm T}_{i}}(V_{\rm g})$, we compute the following root-mean-square-error (RMSE) at each conductance step:
 \begin{equation}
\sigma_{Gi}=\sqrt{\frac{1}{M}\sum_{k}^{M}\left(G_{k}-G_{{\rm fit}_k}\right)^2} .\label{eq:uncerv}
\end{equation}
Here, $M$ is the number of data points in a single conductance step, and $(G_{k}-G_{{\rm fit}_k})$ is the fitting residual.  For a given transition, the RMSE measures how well the data are described by the model we use \cite{Bevington:2003}, and it provides a quantitative estimate for the uncertainty.  The outcome encompasses uncertainties due to noise as well as any physics not taken into account by the Fermi-Dirac step function.  

Equation (\ref{eq:uncerv}) describes the uncertainty in terms of conductance.  However, for spectroscopy purposes, it is preferable to treat the magnetic field as the independent variable, and to express the uncertainty in terms of $B$.  We therefore perform two transformations. The first converts conductance uncertainty into voltage uncertainty for a given transition location $V_{{\rm T}_i}$:
\begin{equation}
\sigma_{V_{{\rm T}_i}}=\sigma_{G_i}\left(\frac{\rmd G}{\rmd V_{\rm g}}\right)^{-1}_{V_{\rm g}=V_{{\rm T}_i}} .
\end{equation}
The second converts voltage uncertainty into magnetic field uncertainty at the transition location $B_{{\rm T}_i}$.  Because of the irregular shape of the of the transition curves, we use the following discretized prescriptions above and below the $i^{\rm th}$ transition, respectively:
\begin{eqnarray}
\sigma_{B_{{\rm T}_i}}^+=B_{{\rm T}_i}(V_{\rm g}+2\sigma_{V_{{\rm T}_i}})-B_{{\rm T}_i}(V_{\rm g}),\label{eq:buncerplus}\\
\sigma_{B_{{\rm T}_i}}^-=B_{{\rm T}_i}(V_{\rm g})-B_{{\rm T}_i}(V_{\rm g}-2\sigma_{V_{{\rm T}_i}}).\label{eq:buncerminus}
\end{eqnarray}

Note that the RMSE does not give a precise measurement of the error in our model.  By comparing the fit to the transition data, however, we find that any additional error can be subsumed into (\ref{eq:buncerplus}) and  (\ref{eq:buncerminus}) through the factors of 2, which have been included.
The resulting expressions for $\sigma_{B_{{\rm T}_i}}^\pm$ provide an upper bound on the uncertainty for any physics left out of our model \cite{McGuire:Thesis}.

\subsection{Energy fitting}\label{sec:routine}
We now perform a second fitting procedure, to extract spectroscopic information from the transition maps  obtained in the previous section.  For a fixed value of $V_{\rm g}$ in figure~\ref{fig:TransBslice}, each conductance transition specifies an equation $E_i=E_{\rm F}$, with $E_i$ given in (\ref{eq:En1}).  The resulting set of (up to) 8 equations is solved simultaneously, obtaining a possible set of conductance transitions, $B_{{\rm fit}_i}(V_{\rm g})$, which depend on the fitting parameters $eV_0$, $\hbar \omega_0$, and any parameters appearing in the valley splitting model.  (Recall that $k=0$.)  In the phenomenological form for the valley splitting, there are 3 fitting parameters, $\Delta_{V_{\rm g}0}$, $\Delta_{B0}$, and $\Delta_{B1}$.  For the theoretical form, there are 2 parameters, $\tilde{\Gamma}$ and $\tilde{A}$.  
By adjusting these parameters, we can minimize the difference between $B_{{\rm fit}_i}$ and $B_{{\rm T}_i}$ by minimizing $\chi^2$, defined as
\begin{equation}
\chi^2=\sum^N_i \frac{1}{\sigma^2_{B_{{\rm T}i}}}(B_{{\rm T}i}-B_{{\rm fit}_i})^2\label{eq:chi2}.
\end{equation}

\begin{figure}[t]
 \begin{center}  \includegraphics[width=5in]{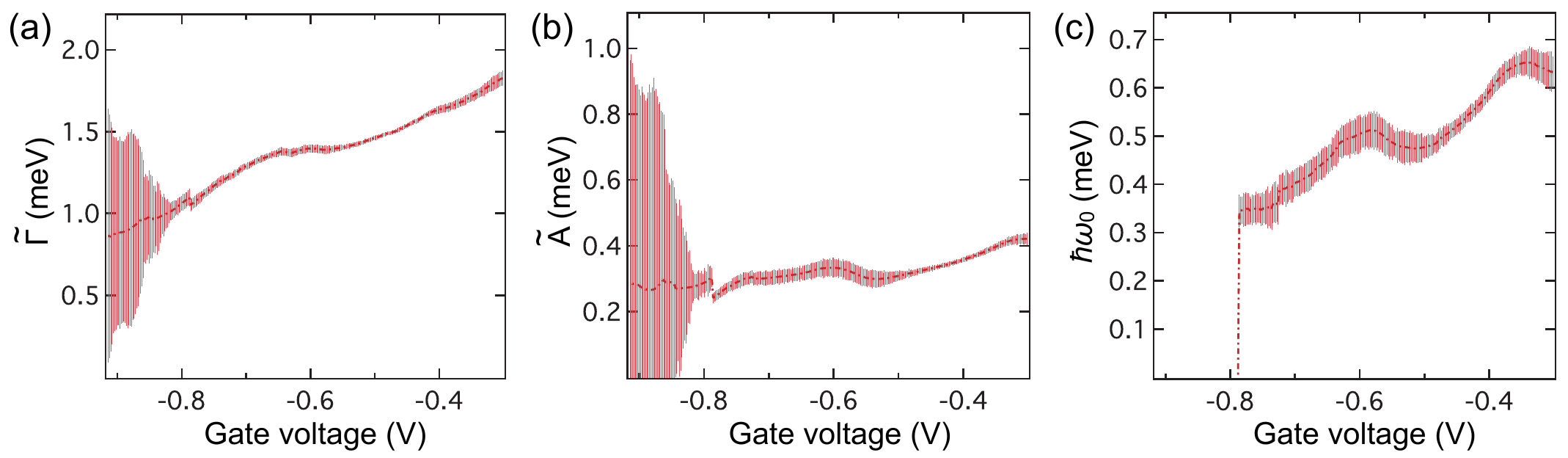}  \end{center}  
 \caption{ \label{fig:NewVSParams}
Fitting results for the valley splitting parameters obtained from the theoretical model, as a function of gate voltage.  Shaded regions show the fitting uncertainty.}
\end{figure}

\subsection{Error propagation}\label{sec:propagate}
The transition uncertainties computed in section~\ref{sec:uncertainty} lead to a range of energy uncertainties in section~\ref{sec:routine}.  To generate error bars for the energy fitting parameters, a Monte Carlo error analysis was implemented, in which the fitting procedure was repeated many times, using randomly generated instances of the transition fields within the uncertainty range.  Since the uncertainty is usually dominated by physical effects left out of the model, there is no particular reason to believe the uncertainty will follow a normal distribution.  We therefore employed an unweighted distribution to generate the transition fields.  The uncertainty that we report for our fitting parameters corresponds to the standard deviation arising from 50 randomized instances.

\subsection{Best fits}\label{sec:bestfits}
Figure~\ref{fig:DNAall} shows experimental conductance transitions, taken from section~\ref{sec:transmain}, together with the energy model fits obtained in section~\ref{sec:routine}.  Results are shown for both the phenomenological and theoretical fitting forms.  The uncertainties shown here are for the transition location, as described in section~\ref{sec:uncertainty}.

 \begin{figure}[t]
\centering
 \begin{center}  \includegraphics[width=4.7in]{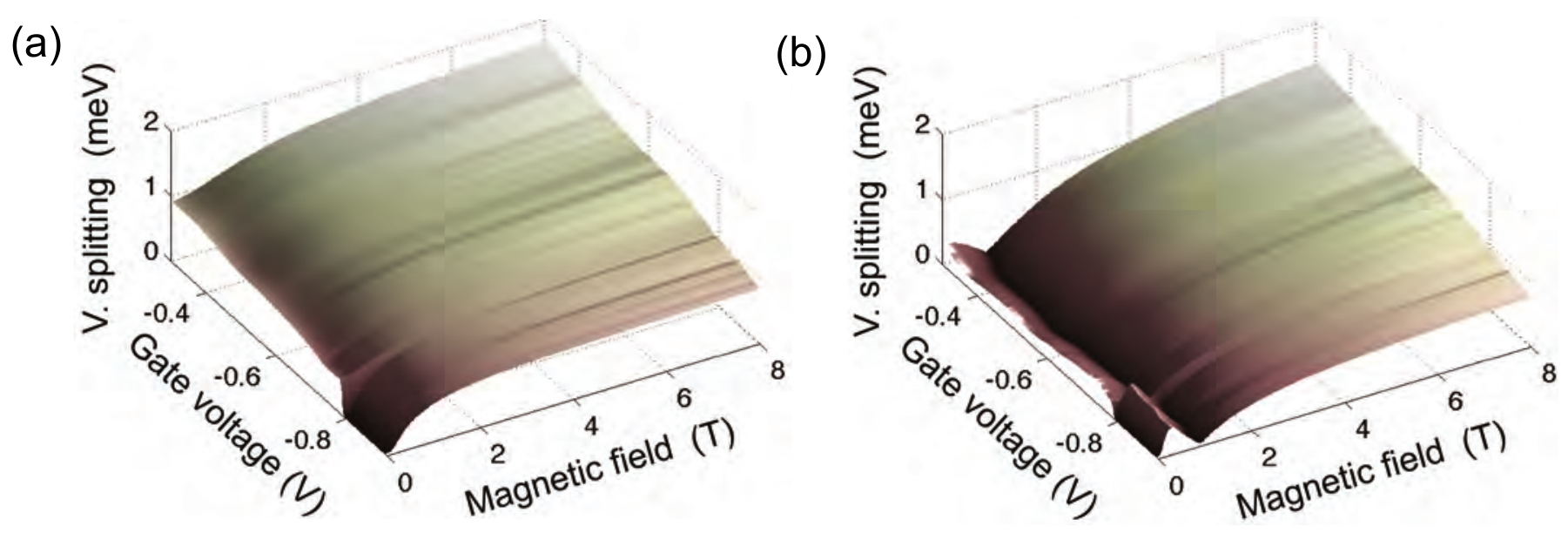}  \end{center}  
 \caption{\label{fig:NewVS3D}
Reconstructed valley splitting for the theoretical fitting form, as a function of magnetic field and gate voltage.  (a) $n=0$ FD level.  (b) $n=1$ FD level.}
\end{figure}

\begin{figure}[t]
 \begin{center}  \includegraphics[width=4.in]{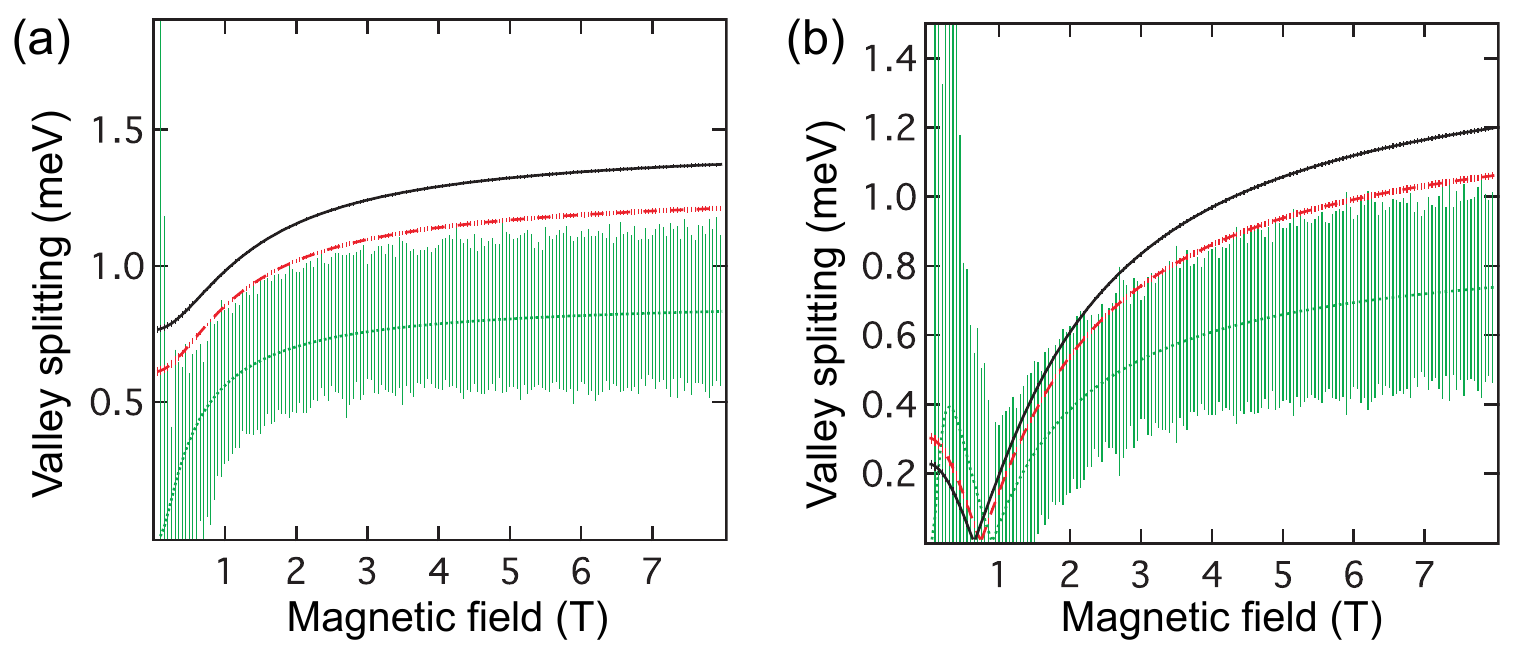}  \end{center}  
 \caption{\label{fig:NewVSslices}
Reconstructed valley splitting for the theoretical fitting form.  Shaded areas indicate the propagated uncertainties.  (a) $n=0$ FD level.  (b) $n=1$ FD level. Solid black lines:  $V_{\rm g}=-0.5$~V.  Dashed red lines:  $V_{\rm g}=-0.7$~V.  Dotted green lines: $V_{\rm g}=-0.9$~V.}
\end{figure}

Note that energy fitting analysis with the theoretical fitting form was restricted to the range $-0.914<V_{\rm g}<-0.3$~V.  Outside this range, there are fewer transitions, and the equations are under-constrained.  The phenomenological fit involves one more fitting parameter, so the equations are under-constrained over a wider range.  Consequently, the procedure is restricted to $-0.8<V_{\rm g}<-0.3$~V.  Note that while the phenomenological fitting form appears to provide a more consistent match to the data in figure~\ref{fig:DNAall}, it also involves a larger number of fitting parameters. 

\begin{figure}[t] 
 \begin{center}  \includegraphics[width=5in]{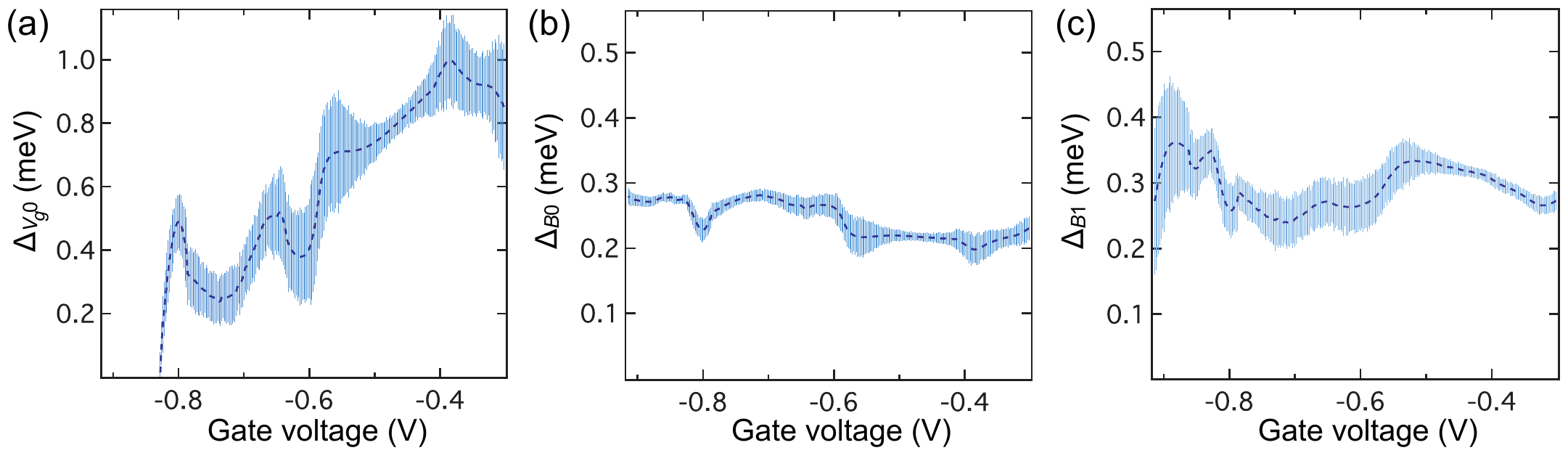}  \end{center}  
 \caption{ \label{fig:OldVSparams}
Fitting results for the valley splitting parameters obtained from the phenomenological fitting form, as a function of gate voltage.  Shaded regions show the fitting uncertainty.}
\end{figure}

\section{Results}\label{sec:results}
Our results for the theoretical fitting parameters $\tilde{\Gamma}$, $\tilde{A}$, and $\hbar \omega_0$ are shown in figure~\ref{fig:NewVSParams}.  Based on these results, we can reconstruct the valley splitting, as shown in figure~\ref{fig:NewVS3D}.  The propagated uncertainties for this analysis were discussed in section~\ref{chap:analysis}.  The computed uncertainties are shown in figure~\ref{fig:NewVSslices} for several representative gate voltages.  

As discussed above, no fitting was performed at low fields ($B<1$~T).  The valley splittings plotted in this range are extrapolations, based on the field-independent fitting parameters obtained at higher fields.  Similarly, the conductance transition data are somewhat sparse at high magnetic fields.  For the $n=0$ level, the valley splitting parameters at magnetic fields $B>4.5$~T were extrapolated from lower fields.  For the $n=1$ level, parameters were extrapolated in the range $B>2.5$~T.

\begin{figure}[t]
 \begin{center}  \includegraphics[width=4.5in]{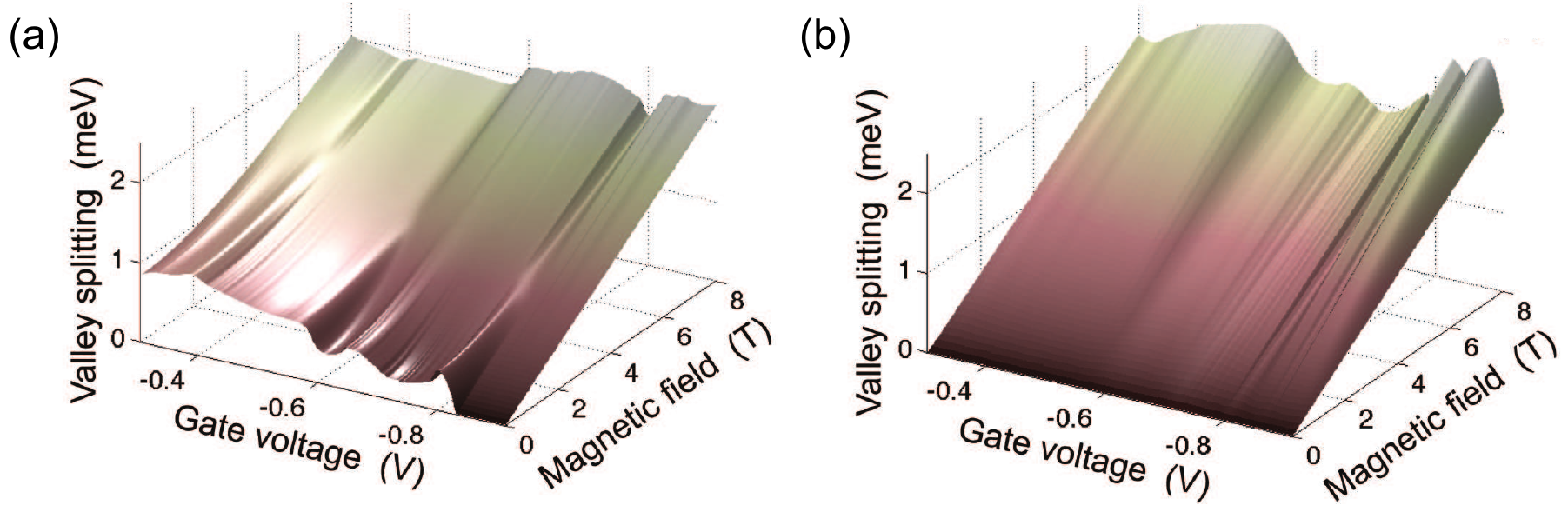}  \end{center}  
 \caption{\label{fig:OldVS3D}
Reconstructed valley splitting for the phenomenological fitting form, as a function of magnetic field and gate voltage.  (a) $n=0$ FD level.  (b) $n=1$ FD level.}
\end{figure}

\begin{figure}[t]
\begin{center}   \includegraphics[width=4.in]{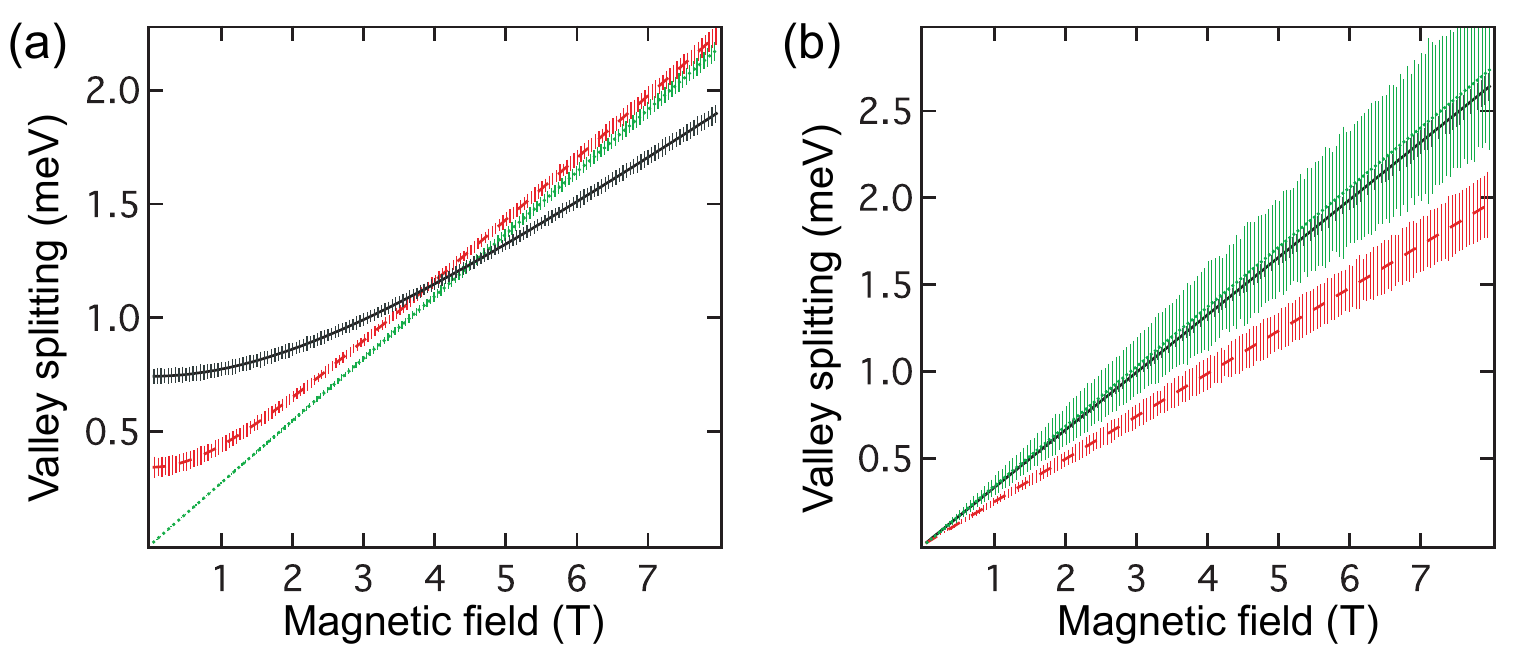} \end{center}
\caption{\label{fig:OldVSslices}
Reconstructed valley splitting for the phenomenological fitting form.  Shaded areas indicate the propagated uncertainties.  (a) $n=0$ FD level.  (b) $n=1$ FD level. Solid black lines:  $V_{\rm g}=-0.5$~V.  Dashed red lines:  $V_{\rm g}=-0.7$~V.  Dotted green lines: $V_{\rm g}=-0.9$~V. }
\end{figure}

In figure~\ref{fig:OldVSparams}, we show results for the phenomenological fitting parameters, $\Delta_{V_{\rm g}0}$, $\Delta_{B0}$, and $\Delta_{B1}$. 
Figure~\ref{fig:OldVS3D} shows the reconstructed valley splitting.  In figure~\ref{fig:OldVSslices} we show the propagated uncertainty for the valley splitting at several fixed gate voltages.   In figures~\ref{fig:OldVS3D} and \ref{fig:OldVSslices}, we also include low and high-field extrapolations of our fitting results, as we did for the theory plots.

The valley splitting results shown in figures~\ref{fig:OldVS3D} and \ref{fig:OldVSslices} agree fairly well with those obtained in Goswami \emph{et al.}~\cite{Goswami:2007p41}.  We attribute the slight differences to the Fermi-Dirac fitting procedure, used here to determine the conductance transitions, and to our restriction of the energy fitting to the range $B>1$~T.  In the voltage range $-0.7<V_{\rm g}<-0.3$~V, the same number of transitions were fit in both analyses and the reconstructed valley splittings were found to differ by less than 0.3~meV.  For $V_{\rm g}<-0.7$~V, the valley splittings reported here differ by up to about 1.0~meV from those in Goswami \emph{et al.}, because a different number of transitions were fit.

\section{Discussion}\label{sec:discussion}
\subsection{Comparison of fitting results}
\begin{figure}
 \begin{center}  \includegraphics[width=5.in]{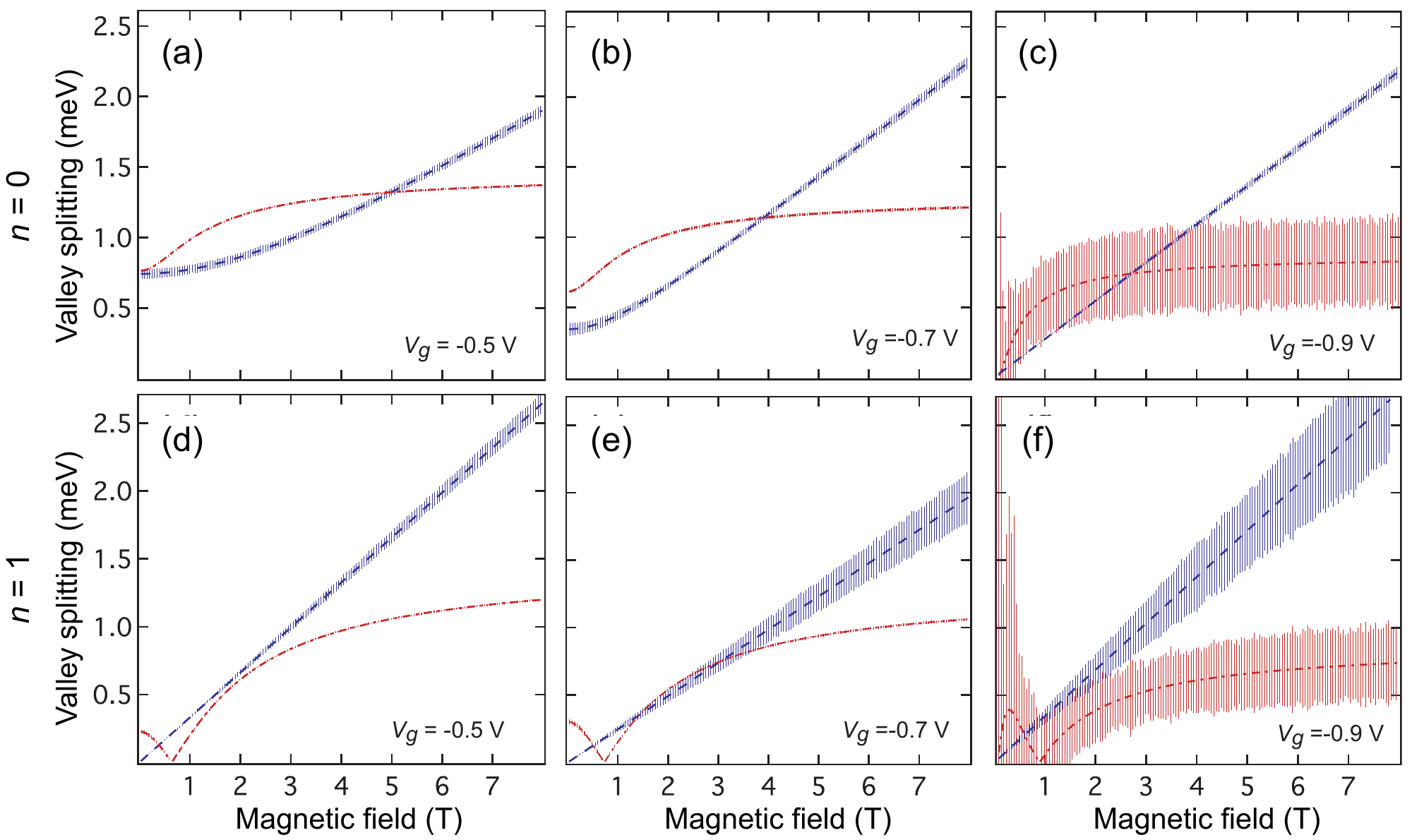}  \end{center}  
  \caption{ \label{fig:compSlices}
  Valley splitting results for the phenomenological fitting form (dashed blue) compared to the theoretical fitting form (dash-dot red), for three gate voltages. The top row shows the valley splitting for the $n=0$ FD level at (a) $V_{\rm g}=-0.5$~V, (b) $V_{\rm g}=-0.7$~V, and (c) $V_{\rm g}=-0.9$~V.  The bottom row shows the valley splitting for the $n=1$ FD level at (d) $V_{\rm g}=-0.5$~V, (e) $V_{\rm g}=-0.7$~V, and (f) $V_{\rm g}=-0.9$~V.  Shaded areas indicate propagated uncertainty.}
 \end{figure}

In the context of quantum computing, we are particularly interested in the overall magnitude of the valley splitting.  It is therefore essential to know whether the results of our analysis depend sensitively on the fitting form that we use.  
In figure~\ref{fig:compSlices}, we compare the reconstructed valley splittings for the phenomenological and theoretical fitting forms.  In the region where the transition data are dense, corresponding to $1< B <4.5$~T for $n=0$ and $1<B<2$~T for $n=1$, we observe that the results do not depend sensitively on the fitting forms, and the two models yield results that differ by no more than 0.5~meV.  This is an important conclusion for our analysis:  the extracted valley splitting is not strongly model dependent in the regime where the data are dense.

For high magnetic fields, the reconstructed valley splittings in figure~\ref{fig:compSlices} exhibit very different behaviors.
The theoretical fitting form saturates while the phenomenological fitting form increases linearly.  We have argued that the extrapolation of the phenomenological form must be incorrect at high field, on physical grounds.  The theoretical form does not suffer from this problem.

It is interesting to note that the low-field extrapolations obtained from the two fitting forms agree in their main features:  a significant zero-field valley splitting for the $n=0$ level, and a relatively small splitting for the $n=1$ level.  
The non-monotonic behavior of the $n=1$ FD level is a conspicuous theoretical feature, and the same feature was observed in a previous numerical analysis \cite{Lee:2006p245302}.  In the present experiment, such non-monotonic behavior occurs in the field range $B<1$~T, where the transitions are not well-resolved, precluding the possibility of confirming such behavior here.

\subsection{Characterizing the quantum well interface} \label{sec:angles}
The theoretical expressions for the valley splitting, obtained in the appendix, incorporate materials parameters describing the substrate miscut angle $\theta$ with respect to [001], and the relative orientation of the QPC device $\varphi$ with respect to [100].  The relative orientation of the QPC to the atomic steps at the interface could not be  determined for this sample; for similar samples, the average orientation was about 45$^\circ$.
By using our transition fitting procedure with the theoretical fitting forms (\ref{eq:D0theory}) and (\ref{eq:D1theory}), it is possible to place some bounds on the values of $\theta$ and $\varphi$ for this sample from the experimental data.

The main obstacle for determining $\theta$ and $\varphi$ lies in (\ref{eq:theory1}), which contains an additional device parameter, $\Delta_{{\rm max}}$, the valley splitting `maximum' for a perfectly flat quantum well interface.  $\Delta_{{\rm max}}$ can be estimated theoretically \cite{Friesen:2007p115318} using the areal density of our 2DEG ($n \simeq 5.72\times 10^{11}$~cm$^{-2}$), giving $\Delta_{{\rm max}} \simeq 0.5$~meV.  We can then estimate $\tilde{\Gamma}$ as follows.  In (\ref{eq:theory1}), we note that $\sin(x)/x$ has a maximum value of 1.  $\tilde{\Gamma}$ should then have a maximum value of 0.3~meV.  Our experimental estimates for $\tilde{\Gamma}$, however, range between 0.9 and 1.8~meV, as shown in figure~\ref{fig:NewVSParams}.  This discrepancy between experiment and theory arises in other 2DEG valley splitting experiments \cite{Takashina:2006p236801,Khrapai:2003p113305}.  The effect is usually attributed to many-body correlations, which are not present in single-electron theories. Indeed, our results are consistent with other recent observations of enhanced valley splitting \cite{Wilde:2005p165429}.  

\begin{figure}[t]
 \begin{center}  \includegraphics[width=3.8in]{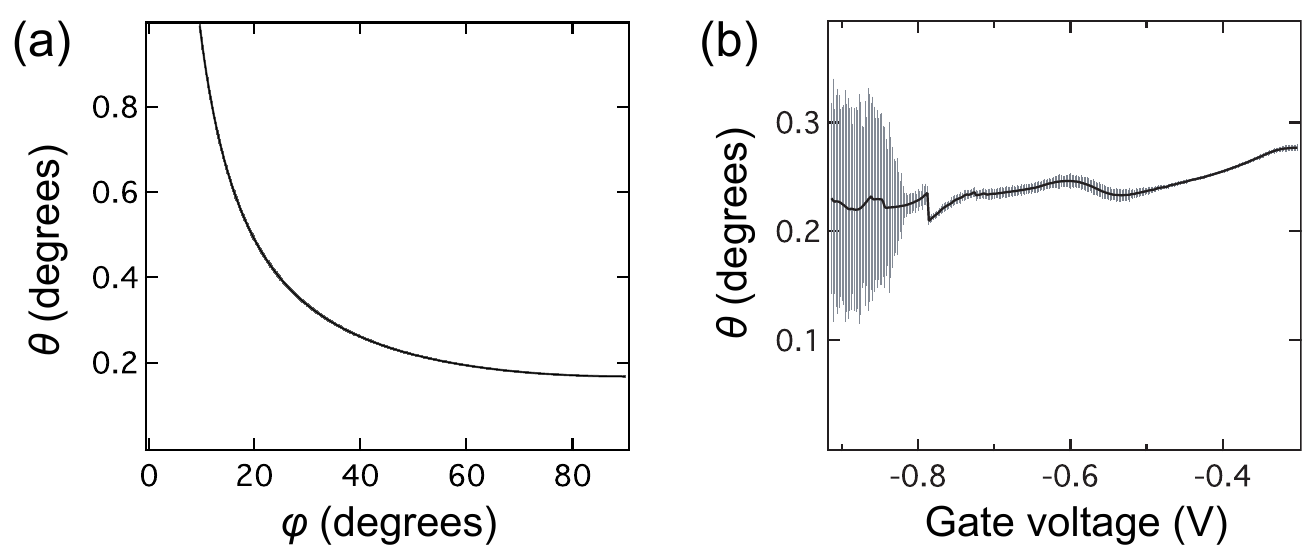}  \end{center}  
 \caption{ \label{fig:Theta}
 (a) Tilt angle $\theta$ as a function of orientation angle $\varphi$, for the case $V_{\rm g}= -0.5$~V. (b) Tilt angle as a function of gate voltage, for the case $\sin^2(\varphi)=1/2$.  Shaded areas indicate uncertainty.}
\end{figure}

Since the value of $\Delta_{{\rm max}}$ is not well known, it becomes impossible to determine $\theta$ and $\varphi$ independently.
However, we can place an approximate bound on $\theta$, using (\ref{eq:Atilde}) and the fitting results for $\tilde{A}$. For example, the dependence of $\theta$ on $\varphi$ is shown in figure~\ref{fig:Theta}.
With the exception of the very smallest rotation angles ($\varphi<20^\circ$), we see that $\theta$ varies slowly in the range $0.15<\theta<0.45^\circ$.  Thus, a characteristic value of $\theta$ is obtained by setting $\sin^2 \varphi = 1/2$ in (\ref{eq:Atilde}).  The result, shown in figure~\ref{fig:Theta}(b), suggests that $\theta \simeq 0.25^\circ$.  

\subsection{Dependence of valley splitting on the Fock-Darwin level}
A main conclusion of this paper, and of several recent papers \cite{Goswami:2007p41,Friesen:2006p202106,Friesen:2007p115318,Kharche:2007p092109,Friesen:2009preprint}, is that in the presence of a miscut substrate or a rough quantum well interface, the observed valley splitting will depend sensitively on the lateral extent of the electron wavefunction.  In a QPC, the lateral extent of the wavefunction depends on both the electrostatic confinement due to the top-gates, and the magnetic confinement from a perpendicular $B$-field.  In figure~\ref{fig:compSlices}, such behavior can be observed clearly for the $n=0$ FD level as a function of magnetic field.  For higher magnetic fields, the magnetic confinement increases and valley splitting is enhanced.  This trend does not depend on which fitting form we use in our analysis.  At low fields, the valley splitting is non-zero, due to electrostatic confinement.

On the other hand, by comparing panels (a)-(c) in figure~\ref{fig:compSlices}, we note that both fitting forms agree that the valley splitting decreases with increasing electrostatic confinement.  This is an experimental fact, which cannot been explained by the theory presented in the appendix.  Here, we attribute this behavior to electron correlation effects.  For weak electrostatic confinement, the electron density in the QPC is relatively high, since there are many electron modes with $k>0$ lying below the Fermi level.  As argued elsewhere, such electron correlations should enhance the valley splitting \cite{Wilde:2005p165429,Takashina:2006p236801,Khrapai:2003p113305}.

The behavior of the $n=1$ FD level is generally consistent with these confinement arguments.  In figure~\ref{fig:compSlices}, similar trends are observed for both $n=0$ and $n=1$.  However, at low fields, the valley splitting behavior is complicated by the nontrivial nodal structure of the $n=1$ wavefunctions, as apparent in the theoretical fitting form.  Generally, one might expect $n=1$ wavefunctions to experience a smaller valley splitting than $n=0$, since the wavefunctions are more extended.  Such behavior is observed in the theoretical fitting form at high fields.  At low fields, however, the nodal structure of the wavefunctions interferes with our expectations.  In fact, the low-field valley splitting in panel (f) is actually larger than panel (c).  It is worth noting that the valley splitting for the $n=1$ FD level can actually be tuned to zero by the application of a magnetic field.

\section{Conclusions}
In this paper, we extracted the valley splitting from conductance measurements in a Si/SiGe QPC using two different fitting forms:  a phenomenological form similar to the one used in \cite{Goswami:2007p41} and a theoretical form developed here.   For both models, the results support the hypothesis that steps at the quantum well interface lead to a dependence of the valley splitting on the spatial extent of the electron wavefunction.  For the lowest transverse mode of the QPC, the valley splitting ranges from about 0.5~meV at zero magnetic field to 1.5~meV at high fields. For the first excited mode, the valley splitting varies non-monotonically and can be tuned from 0 to about 1 meV.

\section*{Acknowledgments}
We would like to thank J. O. Chu of the IBM T. J. Watson Research Center for providing the heterostructure used here. This work was supported by NSA/ARO contract nos.\ W911-NF-04-1-0389 and W911-NF-08-1-0482, by NSF grant nos.\
DMR-0325634,  DMR-0805045, CCF-0523675, and CCF-0523680, and Sandia National Laboratory agreement no.\ 649787.  Sandia is a multiprogram laboratory operated by Sandia Corporation, a Lockheed Martin Company, for the US DOE under contract no.\ DE-AC04-94AL85000.  This research is part of Sandia's Laboratory Directed Research and Development (LDRD) program.

\appendix

\section{Theory of valley splitting in a quantum point contact}\label{sec:VStheory}
\begin{figure}[t] 
 \begin{center}  \includegraphics[width=4in,keepaspectratio]{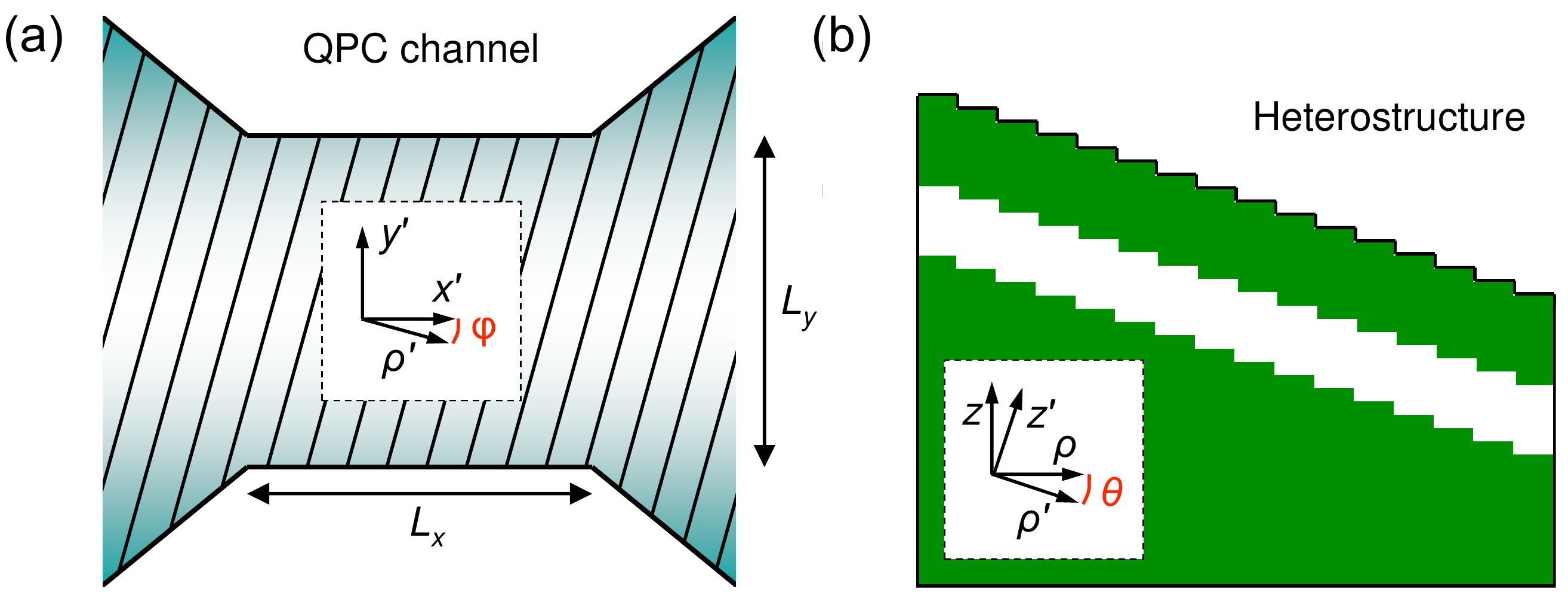}  \end{center}  
\caption{
Schematic QPC geometry, including steps.
(a)  Top view of QPC. $L_x$ and $L_y$ are the length and width of the QPC, respectively. 
Interfacial steps are shown as diagonal lines, rotated from the channel axis by angle $\varphi$.
The downhill direction, perpendicular to the uniform steps, 
corresponds to $\hat{\bm \rho}'$.
(b)  Cross-section view of the steps, with miscut angle $\theta$.  Heterostructure
layers, from bottom to top, correspond to substrate, quantum well, and spacer.  
Doping layer and top-gates are not pictured.}
\label{fig:QPC}
\end{figure}

In this appendix, we develop the valley splitting theory for a single electron confined in QPC geometry, in the presence of a tilted quantum well.  The effective mass envelope function equation for the confined electron is 
\begin{eqnarray}  & & \hspace{-.2in}
\Biggl[ \sum_{n=1}^3 \frac{1}{2m_n} \left( -i\hbar \frac{\partial}{\partial x'_n} 
+e A_n(\bi{r}') \right)^2 \Biggr. 
\label{eq:schrorot}
\\ & & \hspace{0.6in} \Biggl. 
+ V_{\rm QW}(z') +V(x',y')  \Biggr] F(\bi{r}') = E
\, F({\bm r'}) . \nonumber
\end{eqnarray}
Here, the primed coordinate system refers to the plane of the interface, as shown in 
figure~\ref{fig:QPC}.  We have made the usual assumption that the vertical and lateral confinement potentials [$V_{\rm QW}(z') $ and $V(x',y')$, respectively] are separable.

The transverse and lateral 
effective masses are given by $m_1=m_2=m_{\rm t}=0.19\, m_0$ and $m_3=m_{\rm l}=0.92\, m_0$, 
respectively.  Note that the tilted interface induces off-diagonal terms in the inverse effective mass tensor, in the primed coordinate system.  However, for typical tilt angles, these terms comprise less than 3\% of the diagonal elements, and can be treated perturbatively \cite{Friesen:2007p115318}.  In the present work, we ignore them.

Since the atomic steps are much smaller than
any effective mass length scale, we can make the approximation that 
the quantum well potential $V_{\rm QW}(z')$ 
is smoothly tilted (\textit{i.e.}, not step-like).  In the perpendicular field geometry, ${\bm B}=B \hat{\bm z}^\prime$, separation of variables 
leads to solutions of form $F({\bm r}')=F_{xy}(x',y')F_z(z')$, 
where $F_z(z')$ is the quantum well wavefunction and $F_{xy}(x',y')$ is the lateral wavefunction.

We consider the specific QPC geometry shown in figure~\ref{fig:QPC}.  The confinement
along $\hat{\bm y}'$ is modelled as a parabolic potential.  In the $\hat{\bm x}'$
direction, there is no confinement, but we assume periodic boundary conditions
over the distance $L_x$.  The step density is taken to be uniform, with step
orientation given by the angle $\varphi$ relative to the QPC geometry.

We now solve (\ref{eq:schrorot}) by what amounts to a Fock-Darwin (FD) analysis \cite{Fock:1928p446,Darwin:1930p86}.  The zero-field channel width $L_y$ is related to
the oscillator energy $\hbar \omega_0$ as
\begin{equation}
\omega_0 =\frac{\hbar}{m_{\rm t}L_y^2}.
\end{equation}
There is an analogous relation between the cyclotron frequency, $\omega_{\rm c}=e|B|/m_{\rm t}$, and the
magnetic length:
\begin{equation}
\omega_{\rm c} =\frac{\hbar}{m_{\rm t}l_B^2} .
\end{equation}
Working in the primed coordinate system, we then have
\begin{equation}
H=\frac{1}{2m_{\rm t}} \left( -i\hbar {\bm \nabla}'+e{\bm A} \right)^2
+\frac{1}{2}m_{\rm t}\omega_0^2{y'}^2 +V_{\rm QW}(z') .
\end{equation}
The quantum well is perpendicular to $\hat{\bm z}^\prime$, and we define the position of the top interface to be at $z^\prime=0$.  As typical in a 2DEG, we assume a strong modulation doping field so that the electron wavefunction feels only the top interface of the quantum well.  The resulting valley coupling potential is given by \cite{Friesen:2007p115318}
\begin{equation}
V_{\rm v}(z')\simeq v_{\rm v}\delta (z') , \label{eq:Vcoup}
\end{equation}
which we shall treat via perturbation theory.  The coupling strength depends on the conduction band offset at the quantum well, $\Delta E_{\rm c}$, as $v_{\rm v}\simeq 5.4 \times 10^{-11}\Delta E_{\rm c}$.  Here $v_{\rm v}$ is in units of eV$\cdot$m when $\Delta E_{\rm c}$ is in units of eV.  Note that we have used the more accurate estimate of $v_{\rm v}$ from \cite{Friesen:2007p115318}, based on the many-band tight binding model, rather than the two-band model.

As noted above, the problem is separable in the ($x^\prime, y^\prime$) and $z^\prime$ variables. We now solve the ($x^\prime, y^\prime$) problem. For the present geometry, with the magnetic field oriented
perpendicularly to the interface, it is convenient to use the Landau gauge for the vector potential:
${\bm A}=-By'\hat{\bm x}'$.  Except where noted, we henceforth drop the prime notation.
The resulting Hamiltonian, involving the lateral variables, is given by
\begin{equation}
H=-\frac{\hbar^2}{2m_{\rm t}} 
\left( \frac{\partial^2}{\partial x^2} + \frac{\partial^2}{\partial y^2} \right)
+i\hbar\omega_{\rm c}y\frac{\partial}{\partial x}
+\frac{1}{2}m_{\rm t}(\omega_{\rm c}^2 +\omega_0^2) y^2 . \label{eq:H}
\end{equation}

In (\ref{eq:H}), our choice of gauge permits a further separation of variables.
We consider the ansatz $F(x,y) = \exp (\rmi kx)u(y)$, where we have dropped the $xy$ subscript.  The eigenvalue equation for $u(y)$ becomes
\begin{eqnarray}
&& \Biggl[ \frac{\hbar^2k^2}{2m_{\rm t}} \, \frac{\omega_0^2}{\omega_{\rm c}^2+\omega_0^2}
-\frac{\hbar^2}{2m_{\rm t}} \, \frac{\partial^2}{\partial y^2}
\label{eq:Hy} \\ && \hspace{.2in}
+\frac{1}{2}m_{\rm t}(\omega_{\rm c}^2+\omega_0^2)(y-y_k)^2 \Biggr]u(y) = E_{kn}u(y) ,
\nonumber
\end{eqnarray}
where
\begin{equation}
y_k=\frac{\hbar \omega_{\rm c} k}{m_{\rm t}(\omega_{\rm c}^2+\omega_0^2)} . \label{eq:yk}
\end{equation}
In the $x$ direction, the periodic boundary condition leads to the quantization
\begin{equation}
k=\frac{2\pi}{L_x} j,
\end{equation}
where $j$ is an integer.  

Solutions of (\ref{eq:Hy}) are given by
\begin{equation}
E_{kn}=\frac{\hbar^2k^2}{2m_{\rm t}} \, \frac{\omega_0^2}{\omega^2}
+ \left( n+\frac{1}{2} \right) \hbar \omega \hspace{.3in}
(n=0,1,\dots) ,
\end{equation}
and 
\begin{equation}
F_{kn} (x,y)= \frac{\exp (\rmi kx) }{\sqrt{\pi L_x L \, 2^n n! }}
H_n[(y-y_k)/L] \, e^{-(y-y_k)^2/2L^2} , \label{eq:Fkn}
\end{equation}
where
\begin{eqnarray}
\omega^2 &=& \omega_{\rm c}^2+\omega_0^2 , \nonumber \\
L^{-4} &=& l_B^{-4}+L_y^{-4} , \nonumber
\end{eqnarray}
and $H_n(x)$ is a Hermite polynomial \cite{Abramowitz:1972}.  The solutions described above are known as Fock-Darwin levels.  

The valley splitting is now computed from the prescription \cite{Friesen:2007p115318} 
\footnote{Absolute values have been used inside the integrand, to account for complex wavefunctions.}
\begin{equation}
\Delta_{n} = 2\left| \int\! d{r'}^3  \exp (-2\rmi k_0z) 
\left| F_{kn} (r',\phi')F_{z}(z') \right|^2  V_{\rm v}({\bm r'}) \right| \
,   \label{eq:Dv} 
\end{equation}
where we have reverted to primed coordinates since the exponential factor
$ \exp (-2\rmi k_0z)$ (i.e., the valley phase) is defined with respect to the crystallographic (unprimed) axes.
To complete the integral, we need to express the valley phase
in terms of primed coordinates.  In the continuum approximation
(i.e., not treating the steps as discrete), the following relation applies to the interface:
\begin{equation}
z \simeq z'+\theta y' \sin \varphi - \theta x' \cos \varphi .
\end{equation}
Applying (\ref{eq:Vcoup}), we obtain
\begin{eqnarray}
\Delta_{n} &=& \Delta_{{\rm max}} \Biggl| \int_0^{L_x}\! dx \int_{-\infty}^{\infty}\! dy 
\label{eq:Dvkn} \\ && \hspace{.2in} \times
 \exp [2\rmi k_0 \theta (x\cos \varphi -y\sin \varphi )]  \left| F_{kn}(x,y) \right|^2 \Biggr| ,
\nonumber 
\end{eqnarray}
where we have again dropped the prime notation.  

In (\ref{eq:Dvkn}), the parameter $\Delta_{{\rm max}}=2v_{\rm v}F^2_z(0)$ corresponds to the theoretical maximum for the valley splitting in a perfectly flat quantum well.  An estimate for  $\Delta_{{\rm max}}$ requires knowledge of the quantum well wavefunction $F_z(z)$.  The wavefunction can be computed analytically for a 2DEG in the Hartree approximation, as described in \cite{Friesen:2007p115318}.  As usual, we adopt the lowest subband approximation.  Expressing the 2DEG density $n$ in units of $10^{11}{\rm  cm}^{-2}$, we obtain $\Delta_{{\rm max}}\simeq 0.086n$, in units of meV.  However, as described in section~\ref{sec:angles}, this procedure leaves out certain many-body effects, leading to an underestimate of the valley splitting.  In the following analysis, we circumvent this problem by folding $\Delta_{{\rm max}}$ into the valley splitting normalization, since it appears as a prefactor in all our valley splitting expressions.

Defining the scaled valley splitting as
$\tilde{E}_{n} = \Delta_{n}/\Delta_{{\rm max}}$, we can obtain results for several FD levels:
\begin{eqnarray}
\tilde{E}_{0} &=& \Gamma (\theta,\varphi) e^{-b} , \label{eq:E0} \\
\tilde{E}_{1} &=& \Gamma (\theta,\varphi) e^{-b}|1-2b| , \label{eq:E1} \\
\tilde{E}_{2} &=& \Gamma (\theta,\varphi) e^{-b} |1-(2+\sqrt{2})b|\, |1-(2-\sqrt{2})b|,
\hspace{.2in} \label{eq:E2} 
\end{eqnarray}
where we have defined
\begin{equation}
\Gamma (\theta,\varphi) = \frac{1}{\sqrt{\pi}}
\left| \frac{\sin (k_0 L_x \theta \cos \varphi )}
{k_0 L_x \theta \cos \varphi } \right| . \label{eq:Gamma}
\end{equation}
and 
\begin{equation}
b=(k_0 L \theta \sin \varphi )^2\label{eq:b}.
\end{equation}
The geometrical factor $\Gamma (\theta,\varphi)$ oscillates rapidly as a function of
$\theta$ and $\varphi$, reflecting the fact that $\theta$ and $\varphi$ control the
number of steps inside the channel of the QPC.  For the fitting analysis presented in the main text, the equations derived above can be reduced to simple fitting forms, as given in (\ref{eq:D0theory}) and (\ref{eq:D1theory}).

A typical picture of valley splitting in a QPC is obtained by evaluating (\ref{eq:E0})-(\ref{eq:E2}) for several realistic geometries, as shown above in figure~\ref{fig:VSQPC}.  Note that only the $n=0,1$ FD levels are plotted.  Comparing the three panels, we observe that the tilt angle $\vartheta$ plays a strong role in determining the magnetic field scale, since it is directly related to the number of steps covered by the wavefunction.  The width of the QPC plays a strong role at low fields, where the confinement is predominantly electrostatic, rather than magnetic.
At higher magnetic fields, we observe oscillations in the valley splitting caused by the nodal structure of the wavefunctions when $n>0$.  These oscillations were first reported in \cite{Lee:2006p245302}.  Indeed, when we calculate the valley splitting for the device parameters used in \cite{Lee:2006p245302} , we obtain a very similar magnitude and field dependence.  However, the latter work does not take into account the electrostatic confinement of the QPC.

\newpage
\section*{References}
\bibliography{silicon,preprints}

\begin{thebibliography}{10}

\bibitem{Vandersypen:2001p883}
L~M~K Vandersypen, M~Steffen, G~Breyta, C~S Yannoni, M~H Sherwood, and I~L
  Chuang.
\newblock Experimental realization of {S}hor's quantum factoring algorithm
  using nuclear magnetic resonance.
\newblock {\em Nature}, 414(6866):883--887, Jan 2001.

\bibitem{Turchette:1995p4710}
Q~A Turchette, C~J Hood, W~Lange, H~Mabuchi, and H~J Kimble.
\newblock Measurement of conditional phase shifts for quantum logic.
\newblock {\em Phys Rev Lett}, 75(25):4710--4713, Jan 1995.

\bibitem{Sorensen:1999p1971}
A~S\o{}rensen and K~M\o{}lmer.
\newblock Quantum computation with ions in thermal motion.
\newblock {\em Phys Rev Lett}, 82(9):1971--1974, Jan 1999.

\bibitem{Averin:1998p659}
D~V Averin.
\newblock Adiabatic quantum computation with {C}ooper pairs.
\newblock {\em Solid State Comm}, 105(10):659--664, Jan 1998.

\bibitem{Loss:1998p120}
D~Loss and D~P Divincenzo.
\newblock Quantum computation with quantum dots.
\newblock {\em Phys Rev A}, 57(1):120--126, Jan 1998.

\bibitem{Kane:1998p133}
B~E Kane.
\newblock A silicon-based nuclear spin quantum computer.
\newblock {\em Nature}, 393(6681):133--137, Jan 1998.

\bibitem{Elzerman:2004p431}
J~M Elzerman, R~Hanson, L~H~W van Beveren, B~Witkamp, L~M~K Vandersypen, and
  L~P Kouwenhoven.
\newblock Single-shot read-out of an individual electron spin in a quantum dot.
\newblock {\em Nature}, 430:431--435, Jan 2004.

\bibitem{Johnson:2005p925}
A~C Johnson, J~R Petta, J~M Taylor, A~Yacoby, M~D Lukin, C~M Marcus, M~P
  Hanson, and A~C Gossard.
\newblock Triplet-singlet spin relaxation via nuclei in a double quantum dot.
\newblock {\em Nature}, 435:925--928, Jan 2005.

\bibitem{Ciorga:2000p16315}
M~Ciorga, A.~S Sachrajda, P~Hawrylak, C~Gould, P~Zawadzki, S~Jullian, Y~Feng,
  and Z~Wasilewski.
\newblock Addition spectrum of a lateral dot from {C}oulomb and spin-blockade
  spectroscopy.
\newblock {\em Phys Rev B}, 61(24):R16315--R16318, Jun 2000.

\bibitem{Petta:2005p161301}
J~R Petta, A~C Johnson, A~Yacoby, C~M Marcus, M~P Hanson, and A~C Gossard.
\newblock Pulsed-gate measurements of the singlet-triplet relaxation time in a
  two-electron double quantum dot.
\newblock {\em Phys Rev B}, 72:161301, Jan 2005.

\bibitem{Petta:2005p2180}
J~R Petta, A~C Johnson, J~M Taylor, E~A Laird, A~Yacoby, M~D Lukin, C~M Marcus,
  M~P Hanson, and A~C Gossard.
\newblock Coherent manipulation of coupled electron spins in semiconductor
  quantum dots.
\newblock {\em Science}, 309:2180--2184, Jan 2005.

\bibitem{Koppens:2006p766}
F~H~L Koppens, C~Buizert, K~J Tielrooij, I~T Vink, K~C Nowack, T~Meunier, L.~P
  Kouwenhoven, and L~M~K Vandersypen.
\newblock Driven coherent oscillations of a single electron spin in a quantum
  dot.
\newblock {\em Nature}, 442:766--771, Jan 2006.

\bibitem{Nowack:2007p1430}
K~C Nowack, F~H~L Koppens, Y~V Nazarov, and L~M~K Vandersypen.
\newblock Coherent control of a single electron spin with electric fields.
\newblock {\em Science}, 318:1430--1433, Jan 2007.

\bibitem{Koppens:2008p236802}
F~H~L Koppens, K~C Nowack, and L~M~K Vandersypen.
\newblock Spin echo of a single electron spin in a quantum dot.
\newblock {\em Phys. Rev. Lett.}, 100(23):236802, Jun 2008.

\bibitem{Reilly:2008p817}
D~J Reilly, J~M Taylor, J~R Petta, C~M Marcus, M~P Hanson, and A~C Gossard.
\newblock Suppressing spin qubit dephasing by nuclear state preparation.
\newblock {\em Science}, 321(5890):817--821, Aug 2008.

\bibitem{PioroLadriere:2008p776}
M~Pioro-Ladri\`{e}re, T~Obata, Y~Tokura, Y-S Shin, T~Kubo, K~Yoshida,
  T~Taniyama, and S~Tarucha.
\newblock Electrically driven single-electron spin resonance in a slanting
  {Z}eeman field.
\newblock {\em Nat Phys}, 4(10):776--779, Aug 2008.

\bibitem{Barthel:2009p160503}
C~Barthel, D.~J Reilly, C.~M Marcus, M.~P Hanson, and A.~C Gossard.
\newblock Rapid single-shot measurement of a singlet-triplet qubit.
\newblock {\em Phys. Rev. Lett.}, 103(16):160503, Oct 2009.

\bibitem{Simmel:1999p10441}
F~Simmel, D.~A Wharam, M.~A Kastner, and J.~P Kotthaus.
\newblock Statistics of the {C}oulomb-blockade peak spacings of a silicon
  quantum dot.
\newblock {\em Phys Rev B}, 59(16):R10441--R10444, Apr 1999.

\bibitem{Rokhinson:2001p035321}
L.~P Rokhinson, L.~J Guo, S.~Y Chou, and D.~C Tsui.
\newblock Spin transitions in a small {S}i quantum dot.
\newblock {\em Phys Rev B}, 63(3):035321, Jan 2001.

\bibitem{Klein:2004p4047}
L~J Klein, K~A Slinker, J~L Truitt, S~Goswami, K~L~M Lewis, S~N Coppersmith,
  D~W van~der Weide, M~Friesen, R~H Blick, D~E Savage, M~G Lagally, C~Tahan,
  R~Joynt, MA~Eriksson, J~O Chu, J~A Ott, and P~M Mooney.
\newblock Coulomb blockade in a silicon/silicon-germanium two-dimensional
  electron gas quantum dot.
\newblock {\em Appl Phys Lett}, 84:4047--4049, Jan 2004.

\bibitem{Zhong:2005p1143}
Z~Zhong, Y~Fang, W~Lu, and C~M Lieber.
\newblock Coherent single charge transport in molecular-scale silicon
  nanowires.
\newblock {\em Nano Lett}, 5:1143--1146, 2005.

\bibitem{Slinker:2005p246}
K~A Slinker, K~L~M Lewis, C~C Haselby, S~Goswami, L~J Klein, J~O Chu, S~N
  Coppersmith, R~Joynt, R~H Blick, M~Friesen, and M~A Eriksson.
\newblock Quantum dots in {S}i/{S}i{G}e 2{DEG}s with {S}chottky top-gated
  leads.
\newblock {\em New J Phys}, 7:246, Jan 2005.

\bibitem{Sakr:2005p223104}
M~R Sakr, H~W Jiang, E~Yablonovitch, and E~T Croke.
\newblock Fabrication and characterization of electrostatic {S}i/{S}i{G}e
  quantum dots with an integrated read-out channel.
\newblock {\em Appl Phys Lett}, 87:223104, Jan 2005.

\bibitem{Berer:2006p162112}
T~Berer, D~Pachinger, G~Pillwein, M~M\"{u}hlberger, H~Lichtenberger,
  G~Brunthaler, and F~Sch\"{a}ffler.
\newblock Lateral quantum dots in {S}i/{S}i{G}e realized by a {S}chottky
  split-gate technique.
\newblock {\em Appl Phys Lett}, 88:162112, Jan 2006.

\bibitem{Jones:2006p073106}
G~M Jones, B~H Hu, C~H Yang, M~J Yang, Russell Hajdaj, and Gerard Hehein.
\newblock Enhancement-mode metal-oxide-semiconductor single-electron transistor
  on pure silicon.
\newblock {\em Appl Phys Lett}, 89(7):073106, Jan 2006.

\bibitem{Klein:2007p033103}
L~J Klein, D~E Savage, and M~A Eriksson.
\newblock Coulomb blockade and kondo effect in a few-electron
  silicon/silicon-germanium quantum dot.
\newblock {\em Appl Phys Lett}, 90:033103, Jan 2007.

\bibitem{Zimmerman:2007p033507}
N~M Zimmerman, B~J Simonds, A~Fujiwara, Y~Ono, Y~Takahashi, and H~Inokawa.
\newblock Charge offset stability in tunable-barrier {S}i single-electron
  tunneling devices.
\newblock {\em Appl Phys Lett}, 90(3):033507, Jan 2007.

\bibitem{Simmons:2007p213103}
C~B Simmons, M~Thalakulam, N~Shaji, L~J Klein, H~Qin, R~H Blick, D~E Savage,
  M~G Lagally, S~N Coppersmith, and M~A Eriksson.
\newblock Single-electron quantum dot in {S}i/{S}i{G}e with integrated charge
  sensing.
\newblock {\em Appl Phys Lett}, 91:213103, 2007.

\bibitem{Shaji:2008p540}
N~Shaji, C~B Simmons, M~Thalakulam, L~J Klein, H~Qin, H~Luo, D~E Savage, M~G
  Lagally, A~J Rimberg, R~Joynt, M~Friesen, R~H Blick, S~N Coppersmith, and M~A
  Eriksson.
\newblock Spin blockade and lifetime-enhanced transport in a few-electron
  {S}i/{S}i{G}e double quantum dot.
\newblock {\em Nat Phys}, 4(7):540--544, Jun 2008.

\bibitem{Angus:2008p112103}
S~J Angus, A~J Ferguson, A~S Dzurak, and R~G Clark.
\newblock A silicon radio-frequency single electron transistor.
\newblock {\em Appl Phys Lett}, 92:112103, Jan 2008.

\bibitem{Liu:2008p073310}
H~W Liu, T~Fujisawa, Y~Ono, H~Inokawa, A~Fujiwara, K~Takashina, and Y~Hirayama.
\newblock Pauli-spin-blockade transport through a silicon double quantum dot.
\newblock {\em Phys Rev B}, 77:073310, Jan 2008.

\bibitem{Fuhrer:2009p707}
A~Fuhrer, M~F{\"u}chsle, T~C~G Reusch, B~Weber, and M~Y Simmons.
\newblock Atomic-scale, all epitaxial in-plane gated donor quantum dot in
  silicon.
\newblock {\em Nano Lett}, 9(2):707--710, 2009.

\bibitem{Zwanenburg:2009p1071}
F~A Zwanenburg, C~E. W. M~Van Rijmenam, Y~Fang, C~M Lieber, and L~P
  Kouwenhoven.
\newblock Spin states of the first four holes in a silicon nanowire quantum
  dot.
\newblock {\em Nano Lett}, 9(3):1071--1079, Feb 2009.

\bibitem{Simmons:2009p3234}
C~B Simmons, M~Thalakulam, B~M Rosemeyer, B~J~Van Bael, E~K Sackmann, D~E
  Savage, M~G Lagally, R~Joynt, M~Friesen, S~N Coppersmith, and M~A Eriksson.
\newblock Charge sensing and controllable tunnel coupling in a {S}i/{S}i{G}e
  double quantum dot.
\newblock {\em Nano Lett}, 9:3234--3238, 2009.

\bibitem{Nordberg:2009p115331}
E~P Nordberg, G~A~T Eyck, H~L Stalford, R~P Muller, R~W Young, K~Eng, L~A
  Tracy, K~D Childs, J~R Wendt, R~K Grubbs, J~Stevens, M~P Lilly, M~A Eriksson,
  and M~S Carroll.
\newblock Enhancement-mode double-top-gated metal-oxide-semiconductor
  nanostructures with tunable lateral geometry.
\newblock {\em Phys Rev B}, 80:115331, 2009.

\bibitem{Nordberg:2009p202102}
E~P Nordberg, H~L Stalford, R~Young, G~A~T Eyck, K~Eng, L~A Tracy, K~D Childs,
  J~R Wendt, R~K Grubbs, J~Stevens, M~P Lilly, M~A Eriksson, and M~S Carroll.
\newblock Charge sensing in enhancement mode double-top-gated metal-oxide-
  charge sensing in enhancement mode double-top-gated metal-oxide-semiconductor
  quantum dots.
\newblock {\em Appl. Phys. Lett.}, 95:202102, 2009.

\bibitem{Friesen:2003p121301}
M~Friesen, P~Rugheimer, D~E Savage, M~G Lagally, D~W van~der Weide, R~Joynt,
  and M~A Eriksson.
\newblock Practical design and simulation of silicon-based quantum-dot qubits.
\newblock {\em Phys Rev B}, 67:121301, Jan 2003.

\bibitem{Morello:2009p081307}
A~Morello, C.~C Escott, H~Huebl, L.~H.~Willems van Beveren, L.~C.~L Hollenberg,
  D.~N Jamieson, A.~S Dzurak, and R.~G Clark.
\newblock Architecture for high-sensitivity single-shot readout and control of
  the electron spin of individual donors in silicon.
\newblock {\em Phys Rev B}, 80(8):081307(R), 2009.

\bibitem{Feher:1959p1219}
G~Feher.
\newblock Electron spin resonance experiments on donors in silicon. {I}.
  {E}lectronic structure of donors by the electron nuclear double resonance
  technique.
\newblock {\em Phys Rev}, 114(5):1219--1244, Sep 1959.

\bibitem{Wilson:1961p1068}
D~K Wilson and G~Feher.
\newblock Electron spin resonance experiments on donors in silicon. {III}.
  {I}nvestigation of excited states by the application of uniaxial stress and
  their importance in relaxation processes.
\newblock {\em Phys Rev}, 124(4):1068--1083, 1961.

\bibitem{Tyryshkin:2003p193207}
A~M Tyryshkin, S~A Lyon, A~V Astashkin, and A~M Raitsimring.
\newblock Electron spin relaxation times of phosphorus donors in silicon.
\newblock {\em Phys Rev B}, 68(19):193207, Nov 2003.

\bibitem{Tahan:2002p035314}
C~Tahan, M~Friesen, and R~Joynt.
\newblock Decoherence of electron spin qubits in {S}i-based quantum computers.
\newblock {\em Phys Rev B}, 66(3):035314, Jan 2002.

\bibitem{Koiller:2001p027903}
B~Koiller, X~Hu, and S~Das Sarma.
\newblock Exchange in silicon-based quantum computer architecture.
\newblock {\em Phys. Rev. Lett.}, 88(2):027903, Dec 2001.

\bibitem{Boykin:2004p115}
T~B Boykin, G~Klimeck, M~A Eriksson, M~Friesen, S~N Coppersmith, P~von Allmen,
  F~Oyafuso, and S~Lee.
\newblock Valley splitting in strained silicon quantum wells.
\newblock {\em Appl Phys Lett}, 84:115--117, Jan 2004.

\bibitem{Boykin:2004p165325}
T~B Boykin, G~Klimeck, M~Friesen, S~N Coppersmith, P~Von Allmen, F~Oyafuso, and
  S~Lee.
\newblock Valley splitting in low-density quantum-confined heterostructures
  studied using tight-binding models.
\newblock {\em Phys Rev B}, 70:165325, Jan 2004.

\bibitem{Eriksson:2004p133}
M~A Eriksson, M~Friesen, S~N Coppersmith, R~Joynt, L~J Klein, K~A Slinker,
  C~Tahan, P~M Mooney, J~O Chu, and S~J Koester.
\newblock Spin-based quantum dot quantum computing in silicon.
\newblock {\em Quantum Information Processing}, 3(1-5):133--146, Feb 2004.

\bibitem{Weitz:1996p542}
P~Weitz, R~Haug, K~von Klitzing, and F~Sch\"{a}ffler.
\newblock Tilted magnetic field studies of spin- and valley-splittings in
  {S}i/{S}i$_{1-x}${G}e$_x$ heterostructures.
\newblock {\em Surface Science}, 361-362:542--546, Jul 1996.

\bibitem{Koester:1997p384}
S~J Koester, K~Ismail, and J~O Chu.
\newblock Determination of spin- and valley-split energy levels in strained
  {S}i quantum wells.
\newblock {\em Semicond Sci Tech}, 12(4):384--388, Jan 1997.

\bibitem{Schumacher:1998p260}
H~W Schumacher, A~Nauen, U~Zeitler, R~J Haug, P~Weitz, A~G~M Jansen, and
  F~Sch\"{a}ffler.
\newblock Anomalous coincidences between valley split landau levels in a
  {S}i/{S}i{G}e heterostructure.
\newblock {\em Physica B}, 256:260--263, Jan 1998.

\bibitem{Wilde:2005p165429}
M~A Wilde, M~Rhode, C~Heyn, D~Heitmann, D~Grundler, F~Sch{\"a}ffler, and R~J
  Haug.
\newblock Direct measurements of the spin and valley splittings in the
  magnetization of a {S}i/{S}i{G}e quantum well in tilted magnetic fields.
\newblock {\em Phys Rev B}, 72(16):165429, Oct 2005.

\bibitem{Lai:2006p076805}
K~Lai, W~Pan, D~C Tsui, S~Lyon, M~M\"{u}hlberger, and F~Sch\"{a}ffler.
\newblock Intervalley gap anomaly of two-dimensional electrons in silicon.
\newblock {\em Phys. Rev. Lett.}, 96(7):076805, Feb 2006.

\bibitem{Lai:2006p161301}
K~Lai, T~M Lu, W~Pan, D~C Tsui, S~Lyon, J~Liu, Y.~H Xie, M~M\"{u}hlberger, and
  F~Sch\"{a}ffler.
\newblock Valley splitting of {S}i/{S}i$_{1-x}${G}e$_x$ heterostructures in
  tilted magnetic fields.
\newblock {\em Phys Rev B}, 73(16):161301(R), Apr 2006.

\bibitem{Goswami:2007p41}
S~Goswami, K~A Slinker, M~Friesen, L~M McGuire, J~L Truitt, C~Tahan, L~J Klein,
  J~O Chu, P~M Mooney, D~W van~der Weide, R~Joynt, S~N Coppersmith, and M~A
  Eriksson.
\newblock Controllable valley splitting in silicon quantum devices.
\newblock {\em Nat Phys}, 3:41--45, Jan 2007.

\bibitem{Ando:1979p3089}
T~Ando.
\newblock Valley splitting in the silicon inversion layer: Misorientation
  effects.
\newblock {\em Phys Rev B}, 19(6):3089, Oct 1979.

\bibitem{Friesen:2006p202106}
M~Friesen, M~A Eriksson, and S~N Coppersmith.
\newblock Magnetic field dependence of valley splitting in realistic
  {S}i/{S}i{G}e quantum wells.
\newblock {\em Appl Phys Lett}, 89:202106, Jan 2006.

\bibitem{Friesen:2007p115318}
M~Friesen, S~Chutia, C~Tahan, and S~N Coppersmith.
\newblock Valley splitting theory of {S}i{G}e/{S}i/{S}i{G}e quantum wells.
\newblock {\em Phys Rev B}, 75:115318, Jan 2007.

\bibitem{Kharche:2007p092109}
N~Kharche, M~Prada, T~B Boykin, and G~Klimeck.
\newblock Valley splitting in strained silicon quantum wells modeled with
  2$\,^{\circ}$ miscuts, step disorder, and alloy disorder.
\newblock {\em Appl Phys Lett}, 90(9):092109, Jan 2007.

\bibitem{Friesen:2009preprint}
M~Friesen and S~N Coppersmith.
\newblock Theory of valley-orbit coupling in a {S}i/{S}i{G}e quantum dot,
  preprint at http://arxiv.org/abs/0902.0777v1, Feb 2009.

\bibitem{VanWees:1991p12431}
B~J~Van Wees, L~P Kouwenhoven, E~M.~M Willems, C~J P~M Harmans, J~E Mooij, and
  C~T Foxon.
\newblock Quantum ballistic and adiabatic electron transport studied with
  quantum point contacts.
\newblock {\em Phys Rev B}, 43(15):12431--12453, May 1991.

\bibitem{Ismail:1995p1077}
K~Ismail, M~Arafa, K~L Saenger, J~O Chu, and B~S Meyerson.
\newblock Extremely high-electron-mobility in {S}i/{S}i{G}e modulation-doped
  heterostructures.
\newblock {\em Appl Phys Lett}, 66(9):1077--1079, Jan 1995.

\bibitem{Tobben:1995p711}
D~T\"{o}bben, D~A Wharam, G~Absteiter, J~P Kotthaus, and F~Sch\"{a}ffler.
\newblock Ballistic electron transport through a quantum point contact defined
  in a {S}i/{S}i$_{0.7}${G}e$_{0.3}$ heterostructure.
\newblock {\em Semicond Sci Tech}, 10(5):711--714, Jan 1995.

\bibitem{Scappucci:2006p035321}
G~Scappucci, L.~Di Gaspare, E~Giovine, A~Notargiacomo, R~Leoni, and
  F~Evangelisti.
\newblock Conductance quantization in etched {S}i/{S}i{G}e quantum point
  contacts.
\newblock {\em Phys Rev B}, 74(3):035321, Jul 2006.

\bibitem{Wang:1992p12873}
S~L Wang, P~C~Van Son, B~J~Van Wees, and T~M Klapwijk.
\newblock Quantum conductance of point contacts in {S}i inversion layers.
\newblock {\em Phys Rev B}, 46:12873--12876, Jan 1992.

\bibitem{Schaffler:1997p1515}
F~Sch\"{a}ffler.
\newblock High-mobility {S}i and {G}e structures.
\newblock {\em Semicond Sci Tech}, 12(12):1515--1549, Jan 1997.

\bibitem{Ando:1982p437}
T~Ando, A~B Fowler, and F~Stern.
\newblock Electronic properties of two-dimensional systems.
\newblock {\em Rev Mod Phys}, 54(2):437--672, Jan 1982.

\bibitem{Sham:1979p734}
L~J Sham and M~Nakayama.
\newblock Effective-mass approximation in the present of an interface.
\newblock {\em Phys Rev B}, 20(2):734--747, Jan 1979.

\bibitem{Ohkawa:1977p907}
F~J Ohkawa and Y~Uemura.
\newblock Theory of valley splitting in an n-channel (100) inversion layer of
  {S}i {I}. {F}ormulation by extended zone effective mass theory.
\newblock {\em J. Phys. Soc. Jpn.}, 43(3):907--916, Apr 1977.

\bibitem{Ohkawa:1978p69}
F~J Ohkawa.
\newblock Electric break-through in an inversion layer: Exactly solvable model.
\newblock {\em Solid State Commun}, 26(2):69--71, 1978.

\bibitem{Nestoklon:2008p155328}
M~O Nestoklon, E~L Ivchenko, J-M Jancu, and P~Voisin.
\newblock Electric field effect on electron spin splitting in {S}i{G}e/{S}i
  quantum wells.
\newblock {\em Phys Rev B}, 77(15):155328, Jan 2008.

\bibitem{Takashina:2006p236801}
K~Takashina, Y~Ono, A~Fujiwara, Y~Takahashi, and Y~Hirayama.
\newblock Valley polarization in {S}i(100) at zero magnetic field.
\newblock {\em Phys. Rev. Lett.}, 96(23):236801, Jun 2006.

\bibitem{Khrapai:2003p113305}
V~S Khrapai, A~A Shashkin, and V~T Dolgopolov.
\newblock Strong enhancement of the valley splitting in a two-dimensional
  electron system in silicon.
\newblock {\em Phys Rev B}, 67(11):113305, Mar 2003.

\bibitem{Bevington:2003}
P~R Bevington and D~K Robinson.
\newblock {\em Data Reduction and Error Analysis for the Physical Data
  Reduction and Error Analysis for the Physical Sciences}.
\newblock Mcgraw-Hill, Boston, 3rd edition, 2003.

\bibitem{McGuire:Thesis}
L~M McGuire.
\newblock {\em Valleys in a {S}i/{S}i{G}e Quantum Point Contact}.
\newblock PhD thesis, University of Wisconsin-Madison, 2008.

\bibitem{Beenakker:1991p1}
C~W~J Beenakker and H~van Houten.
\newblock Quantum transport in semiconductor nanostructures.
\newblock In H~Ehrenreich and D~Turnbull, editors, {\em Solid State Physics:
  Advances in Research and Applications}, volume~44. Academic Press, New York,
  1991.

\bibitem{Reilly:2001p121311}
D~J Reilly, G~R Facer, A~S Dzurak, B~E Kane, R~G Clark, P~J Stiles, R~G Clark,
  A~R Hamilton, J~L O'Brien, and N~E Lumpkin.
\newblock Many-body spin-related phenomena in ultra low-disorder quantum wires.
\newblock {\em Phys Rev B}, 63(12):121311, Mar 2001.

\bibitem{Lee:2006p245302}
S~Lee and P~von Allmen.
\newblock Magnetic-field dependence of valley splitting in {S}i quantum wells
  grown on tilted {S}i{G}e substrates.
\newblock {\em Phys Rev B}, 74(24):245302, Dec 2006.

\bibitem{Fock:1928p446}
V~Fock.
\newblock Bemerkung zur quantelung des harmonischen oszillators im magnetfeld.
\newblock {\em Z Phys}, 47:446--448, 1928.

\bibitem{Darwin:1930p86}
C~G Darwin.
\newblock The diamagnetism of the free electron.
\newblock {\em Proc. Cambridge Philos. Soc.}, 27:86--90, 1930.

\bibitem{Abramowitz:1972}
M~Abramowitz and I~A Stegun, editors.
\newblock {\em Handbook of Mathematical Functions}.
\newblock Dover, New York, 1972.

\end{thebibliography}

\end{document}